\documentclass[a4paper,english]{aa}
\usepackage{times}
\usepackage[varg]{txfonts}
\usepackage[T1]{fontenc}
\usepackage{amsmath}
\usepackage{graphicx}
\usepackage{amssymb}
\usepackage{natbib}
\usepackage{babel}
\usepackage{xspace}
\bibpunct{(}{)}{;}{a}{}{,}

\newcommand{\mc}[3]{\multicolumn{#1}{#2}{#3}}
\newcommand{\mrm}{\mathrm}
\newcommand{\tfm}{\tablefootmark}
\newcommand{\tft}[2]{\tablefoottext{#1}{#2}}
\defcitealias{khea12}{Paper~I}
\defcitealias{mwscat}{Paper~II}
\defcitealias{newari}{Paper~III}
\defcitealias{nchpm}{Paper~IV}
\defcitealias{mwscint}{Paper~V}
\begin{document}

\title{Global survey of star clusters in the Milky Way}

\subtitle{VI. Age distribution and cluster formation history}

\author{A.E.~Piskunov   \inst{1,2} \and
        A.~Just         \inst{3}   \and 
        N.V.~Kharchenko \inst{1,4} \and
        P.~Berczik      \inst{3,4} \and
        R.-D.~Scholz    \inst{5}   \and
        S. Reffert      \inst{1}   \and
        S.X.~Yen          \inst{1}
        }

\offprints{R.-D.~Scholz}

\institute{
Zentrum f\"ur Astronomie der Universit\"at
Heidelberg, Landessternwarte, K\"{o}nigstuhl 12, 69117 Heidelberg, Germany
\and
Institute of Astronomy of the Russian Acad. Sci., 48 Pyatnitskaya Str., 109017
Moscow, Russia
\and
Zentrum f\"ur Astronomie der Universit\"at
Heidelberg, Astronomisches Rechen-Institut, M\"{o}nchhofstra\ss{}e 12-14, 69120 Heidelberg, Germany
\and
Main Astronomical Observatory, 27 Academica Zabolotnogo Str., 03143 Kiev,
Ukraine
\and
Leibniz-Institut f\"ur Astrophysik Potsdam (AIP), An der Sternwarte 16, 14482
Potsdam, Germany\\
email: rdscholz@aip.de
}

\date{Received 22 November 2017 / Accepted 19 February 2018}

\abstract 
%context
{The all-sky Milky Way Star Clusters (MWSC) survey provides uniform and precise ages, along with other relevant parameters, for a wide variety of clusters in the extended Solar Neighbourhood.} 
%aims 
{In this study we construct the cluster age distribution, investigate its spatial variations, and discuss constraints on cluster formation scenarios of the Galactic disk during the last 5 Gyrs.} 
%methods 
{Due to the spatial extent of the MWSC, we consider spatial variations of the age distribution along galactocentric radius $R_G$, and along $Z$-axis. For the analysis of the age distribution we use 2242 clusters, which all lie within roughly 2.5 kpc of the Sun. To connect the observed age distribution to the cluster formation history we build an analytical model based on simple assumptions on the cluster initial mass function and on the cluster mass-lifetime relation, fit it to the observations, and determine the parameters of the cluster formation law.}
%results 
{Comparison with the literature shows that earlier results strongly underestimated the number of evolved clusters with ages $t\gtrsim 100$ Myr. Recent studies based on all-sky catalogues agree better with our data, but still lack the oldest clusters with ages $t\gtrsim 1$ Gyr. We do not observe a strong variation in the age distribution along $R_G$, though we find an enhanced fraction of older clusters ($t>1$ Gyr) in the inner disk. In contrast, the distribution strongly varies along $Z$. The high altitude distribution practically does not contain clusters with $t<1$ Gyr.  With simple assumptions on the cluster formation history, the cluster initial mass function and the cluster lifetime we can reproduce the observations. The cluster formation rate and the cluster lifetime are strongly degenerate, which does not allow us to disentangle different formation scenarios. In all cases the cluster formation rate is strongly declining with time, and the cluster initial mass function is very shallow at the high mass end.} 
%conclusions 
{}


\titlerunning{MWSC VI. Age distribtion \& cluster formation history}

\maketitle

\section{Introduction}\label{sec:intro}

Galactic disk evolution implies temporal variations of the disk population and its constituents. Field stars along with star clusters represent the typical disk population, and have been the subjects of investigation into the evolutionary processes in the Milky Way for a long time. Our understanding of disk evolution is largely based on stellar data, coming from studies of the kinematics, abundances and ages of F-K dwarfs \citep[see e.g.][]{jujah10}.
Aside from the advantage of the long lifetimes of late-G and K dwarfs, which exceed the age of the disk, using stars leads to a number of disadvantages as they represent the very local disk only up to a few hundred parsecs, and their ages and other evolutionary parameters are of lower accuracy than those of star clusters. On the other hand, the typical lifetime of open clusters is lower than the age of the disk, exceeding it only for initially very massive clusters. This raises a demand for studies of old open clusters, which provide insights into the early epochs of the Milky Way disk's formation and evolution.

Thus the investigation of star cluster ages, in conjunction with either the dissolution history of star clusters or with the star formation history of the Milky Way, is of great interest. This is largely connected to the advances in open clusters observations, dating, and compilations of representative samples of open clusters and collection of their data into all-sky catalogues. One can mention for example studies of \citet{wiel71}, \citet{pama86}, \citet{janes88}, \citet{bcd91}, and \citet{clupop} who constructed the local cluster age distributions, and estimated present cluster formation rates (in the range 0.10-0.45 Myr$^{-1}$ kpc$^{-2}$), and typical lifetimes on the order of $100-250$ Myr. \citet{lamea} and \citet{lamgi06} proposed a model to explain the local distribution of open clusters with age. \citet{mora13} fitted this model to their constructed age distribution of clusters observed towards the Galactic center.   

In this paper, we use data from the Milky Way Star Clusters catalog MWSC \citep[][hereafter \citetalias{khea12}]{khea12}, to study cluster ages. Within the MWSC project, \citet[][\citetalias{mwscat}]{mwscat} determined various cluster parameters for 2859 clusters known in the literature, whereas \citet[][\citetalias{newari}]{newari} and \citet[][\citetalias{nchpm}]{nchpm} added 202 newly-discovered open clusters and associations. The full MWSC sample includes clusters with heliocentric distances up to 15\,kpc, with the mode at 2.4\,kpc. The ages and distances in the MWSC survey are based on cluster members with NIR photometric data from the 2MASS catalogue \citep{cat2MASS} that were fitted to the newest Padova isochrones.

The basic purpose of our study is to derive the unbiased age distribution of MWSC clusters in the wider Solar Neighbourhood, and to study its variations within a spatial completeness zone extending between the Sagittarius and Perseus spiral arms. Specifically we aim at understanding if the age distributions from the arm areas differ from that in the inter-arm region. We will also fit a simple analytical model of cluster formation in the Galactic disk to the observations, in order to draw conclusions on the consistency of our main assumptions with the temporal variations of the cluster formation rate, and with the main components of the model (cluster initial mass function and a relation between cluster lifetime and the clusters' initial mass) used for model construction. 

The outline of the paper is as follows: Sect.~\ref{sec:data} gives a short overview of the input data and  the cluster parameters obtained within the MWSC survey. In Sect.~\ref{sec:genage} we describe the working samples of data, a method of construction of the unbiased age distribution of the local clusters, and derived general results. In Sect.~\ref{sec:spvar} we consider spatial variations of the cluster age distribution in the Galactic plane and along the $Z$-axis. In Sect.~\ref{sec:cluhis} we construct a simple analytical model describing the evolution of open clusters during the last 5 Gyrs, fit it to the observed age distribution of local clusters and draw conclusions on the details of cluster formation. Sect.~\ref{sec:conc} summarises our results.

\section{The sample and the data on cluster ages}\label{sec:data}

In order to build the age distribution of a given set of objects, one has to select a complete and unbiased list of the objects, which also has to be uniformly dated by an accurate and unbiased method.

For galactic star clusters, the MWSC survey is an ideal source, as it suits the requirements mentioned above. It provides a comprehensive sample of star clusters together with a number of well-determined parameters based on uniform photometric and kinematic stellar data gathered from the all-sky catalogues 2MASS \citep{cat2MASS} and PPMXL \citep{ppmxl}. A merger of these catalogues, 2MAst \citepalias[see][for a description of the 2MAst construction]{khea12}, was used to verify clusters from an input list, find new clusters and determine cluster parameters in astrometric and photometric systems that are homogeneous over the whole sky. The full MWSC sample contains 3208 objects: 3061 open and 147 globular clusters. In this study, we concentrate on the subset of open clusters. For all stars within the cluster areas, cluster membership probabilities were determined by using kinematic (proper motions) and photometric (colour magnitude diagrams) selection criteria. The procedure is described in \citetalias{khea12}, and the results are published in the MWSC catalogues \citepalias{mwscat,newari,nchpm}. 

\begin{figure}[t]
   \centering
\includegraphics[width=0.99\hsize,clip=]{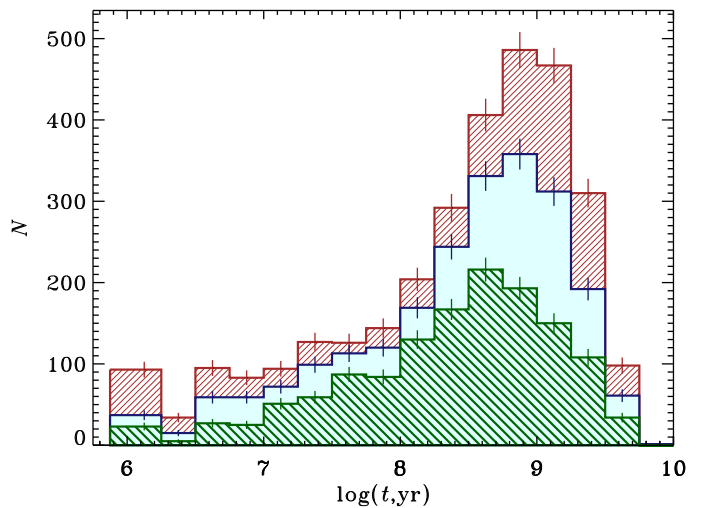}
\caption{Distributions of MWSC clusters with age. The total sample is shown with a background (brown) hatched histogram, clusters within individual completeness limits are shown with an intermediate (blue) filled histogram, and those within the general completeness circle are shown with a foreground (green) back-hatched histogram. The vertical bars show Poisson errors.
}
\label{fig:his_nage}
\end{figure}

The cluster ages are based on uniform cluster membership and present-day isochrones including both pre- and post-MS evolutionary stages. The ages were determined by fitting cluster member CMDs to isochrones computed using the Padova on-line server CMD2.2\footnote{http://stev.oapd.inaf.it/cgi-bin/cmd} \citepalias[for more details see][]{khea12,mwscat}. At metallicities typical of the Galactic disk, the isochrones only weakly depend on the metallicity. Therefore, only a single set of isochrones with solar metallicity was used in the determinations of the MWSC survey. Simultaneously with cluster ages, the respective distances and reddening values were also determined from the isochrone fitting. In the result, all MWSC objects were provided with the homogeneous data necessary for this study. As shown in a comparison with the literature \citepalias[see][]{mwscat}, our ages are typically accurate within 10\% for clusters older than $\log t=8.2$ (with age $t$ in years and which according to Fig.\ref{fig:his_nage} comprises more than 60\% of the total survey) and within 25\% for younger clusters. The distances are accurate within 11\%.

%----------------------------------------------------------------------------%
%-------------------------------------------------------------------------

In total we have age determinations for 3061 star clusters and cluster-like objects (compact associations, regular, embedded, remnant and moving clusters), which we call hereafter open clusters for simplicity. The sample includes effectively all clusters previously known from the literature with the addition of 202 new clusters.  Ages of MWSC clusters cover a considerable fraction of the age of the Galactic disk, spanning between $\log t=6$ and $\log t=9.78$. The cluster sample also covers a wide range of galactocentric distances $R_G$, from the Galactic centre to the outskirts of the Galactic disk at about $R_G=15$ kpc. The completeness zone (where we know virtually all clusters) has a radius of about 2 kpc, reaching the Sagittarius and Perseus spiral arms. 

However, the data completeness is not uniform. The statistics of extremely young and extremely old clusters is still insufficiently known. For example, the density of older clusters in the Solar Neighbourhood is lower than in the outer regions, which implies that a few tens of old clusters within about 1 kpc from the Sun \citepalias{mwscat} are missing. Also the number of the youngest clusters is still uncertain, since many of them may still be obscured by heavy gas-dust clouds in star formation sites. It should also be noted that the MWSC pipeline could not be applied to the nearest clusters, the Ursa Major moving cluster and the Hyades, where 3D motions should be used for member selection. Therefore, these two well-known clusters are not included in our sample.

At the youngest ages, our star cluster sample could be biased by the pre-main sequence isochrones, which are less secure than at the MS- and post-MS stages. The age edge effect should also be considered: the used set of isochrones had a lower age limit of $\log t=6.0$, and no ages younger than this could be determined. Thus, this value was artificially assigned to potentially younger clusters. Furthermore, the quality of the determined MWSC parameters strongly depends on the distance to the cluster. At larger distances, due to a fixed faint limit of the apparent magnitude of the survey, one can only observe the tip of the cluster MS/RG-branches, and so the accuracy of the parameters diminishes.

\section{Cluster age distribution}\label{sec:genage}

The raw age distribution of 3061 MWSC clusters is shown in Fig.~\ref{fig:his_nage} as brown hatched histogram. One can see that it is dominated by old clusters with age $\log t= 8.5...9.5$. The peak is partly due to the logarithmic age scale and partly due to the NIR nature of the survey: the clusters with red giants tend to be brighter than their younger counterparts, are observed at larger distances, and thus are more numerous at the limiting magnitude of the survey, as they are collected from larger areas of the disk. 

\subsection{Data completeness issue}\label{sec:datcompl}

As cluster counts show, the MWSC can be classified as a magnitude-limited sample \citep[see][\citetalias{mwscint} for details]{mwscint}.  The surface density profile for such a sample can be represented schematically by a flat inner area, where the data incompleteness is low/negligible, and by a long outer tail of gradually decreasing density, which is biased by the survey incompleteness at faint magnitudes. The incompleteness can be quantified in statistical sense as a measure of the decrease of the observed surface density with respect to the averaged local density \citep[see e.g.][]{mora13}. Note, that as a measure of the distance we use hereafter a Galactic plane projection $d_{xy}$ of solarcentric distance $d$. The radius of the flat area $\hat{d}_{xy}$ is then called the completeness limit of the survey. Once it is established, the bias-free statistics is gathered within the completeness limit.

This approach (which we call hereafter single-limit approach) is attractive due to its simplicity and was commonly used starting from the pioneering work of \citet{wiel71}, but is inherent in a bias for objects which are absolutely fainter or brighter than the clusters typical of the given sample. For example, when applying the single-limit approach to faint objects, which must be observed near to the Sun only, one underestimates their density, when one divides their counts by the completeness area defined by the common completeness limit. In contrast, since the typical distance to bright objects may exceed the completeness limit, one can lose them from the statistics at all. This is especially important since for a NIR survey (including the MWSC, where this statement is supported by direct statistics) the brightest objects are as a rule the oldest ones, and their loss leads to a bias in the early history of the disk. Therefore, to avoid important biases, which might affect the end-distribution, we decided to abandon the single-limit approach and apply instead a strategy used for the stellar luminosity function construction, which collects stars of different absolute magnitudes from proportionally extended completeness areas. We refer to this approach as  variable completeness limit concept. Note, that this approach represents a development of the single-limit approach, where single completeness zones are prescribed to objects from some narrow absolute magnitude interval. 

This strategy became possible since we have determined in \citetalias{mwscint} for all MWSC clusters their integrated NIR magnitudes, and have built a magnitude-dependent completeness distance scale, and its relation to galactic longitude. For the absolute integrated magnitude in the $K_S$ passband $I(M_{K_S})$ this relation averaged over galactic longitudes can be written as
\begin{equation}
\hat{d}_{xy} = p - q \times I(M_{K_S})\,,         \label{eq:dciksrel}
\end{equation}
with $p=0.80\pm0.05$, and $q=0.42\pm0.02$, where $\hat{d}_{xy}$ is in kpc.

As we determined earlier applying both approaches in \citetalias{mwscat} and in \citetalias{mwscint}, the MWSC is generally complete within about 1.8-2.2 kpc from the Sun (we adopt hereafter the lower limit of this interval), where about half of all MWSC open clusters are located (green back-hatched histogram in Fig.~\ref{fig:his_nage}). Their age-distribution is still similar to the total one, though the contrast between the peak of older clusters and younger ones becomes lower (the ratios between values at the maxima and shoulders at about $\log t =7.9$ are equal to 3.4 and 2.4 respectively). If one adopts a more flexible variable completeness limit the ratio equal to 3.0 falls in between (see also blue histogram in Fig.~\ref{fig:his_nage}).  As seen in Fig.~\ref{fig:his_nage} this approach does not change the shape of the distribution strongly, but involves in the statistics almost 40 percent more clusters (2242 objects), which makes the completeness sample more representative and allows us to look at larger distances from the Sun. 

%----------------------------------------------------------------------------%
\begin{figure}[t]
   \centering
\includegraphics[width=0.99\hsize,clip=]{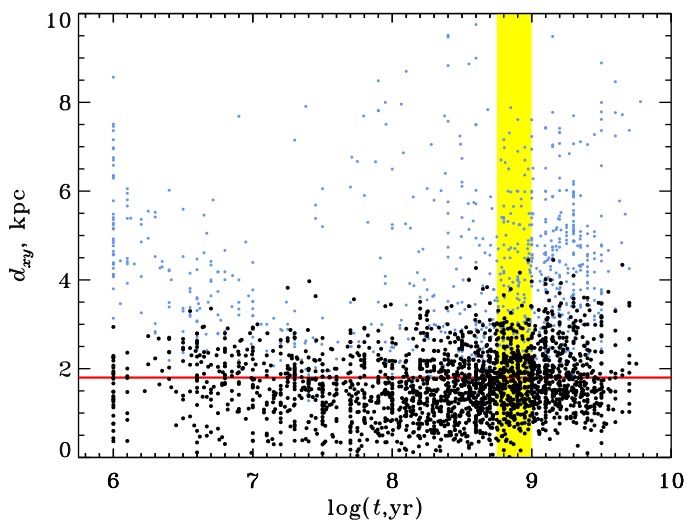}
\caption{Distribution of cluster distances with age. All clusters are shown with light (blue) dots, while clusters located within individual completeness limits computed along Eq.~(\ref{eq:dciksrel}) are shown with black dots. The general completeness limit for the MWSC-sample is given by the horizontal (red) line. A vertical yellow stripe is given for illustration and indicates an arbitrary $\log t$-box with clusters of the two kinds falling in it.
}
\label{fig:dist_age}
\end{figure}
%------------------------------------------~\ref{fig:his_nage}-------------------------------

In Fig.~\ref{fig:dist_age} we compare both completeness approaches in the $d_{xy}$ vs. $\log t$ diagram, which has a distinctive U-shape configuration. It is clear that the most distant clusters belong to either the youngest ($\log t < 7$) or the oldest age group ($\log t \gtrsim 8$). The single completeness limit (despite only including about half of the total objects) seems to be more selective (higher impact of non-regular lower bound in $d_{xy}$, especially strong at youngest and oldest ages). In the alternative case one can extend the size of the completeness area almost by a factor of two (especially for older ages, see Fig.~\ref{fig:dist_age}). On average, the individual completeness approach allows us to expand the total completeness area to about 3 kpc. However, this causes the upper limit of the completeness zone to become non-uniform with respect to age: at $\log t>8.5$  the completeness distance reaches for the intrinsically brightest clusters $d_{xy}=4.5$ kpc. An additional bias is introduced by the inhomogeneous distribution of MWSC clusters over the sky, which is related to the patchy distribution of interstellar extinction and in particular strongly discriminates against clusters of the Pleiades type (those deprived of bright NIR red giants). This results in the lowered cluster density in the $d_{xy}$ vs. $\log t$-diagram at $\log t < 6.6$ for young clusters associated with dust clouds, and $\log t\sim 7.4-8.3$ for screened star clusters (preferentially red giant-deficient). This deficiency should be remembered as a source of bias in the built age distribution.

To the end of Sec.~\ref{sec:genage} we will use both completeness approaches in order to ensure that they give similar results in the limiting case of the local clusters, which is also important for comparison with the literature, where almost exclusively the single completeness limit concept is used.

%----------------------------------------------------------------------------%
\begin{figure}[t]
   \centering
\includegraphics[width=0.99\hsize,clip=]{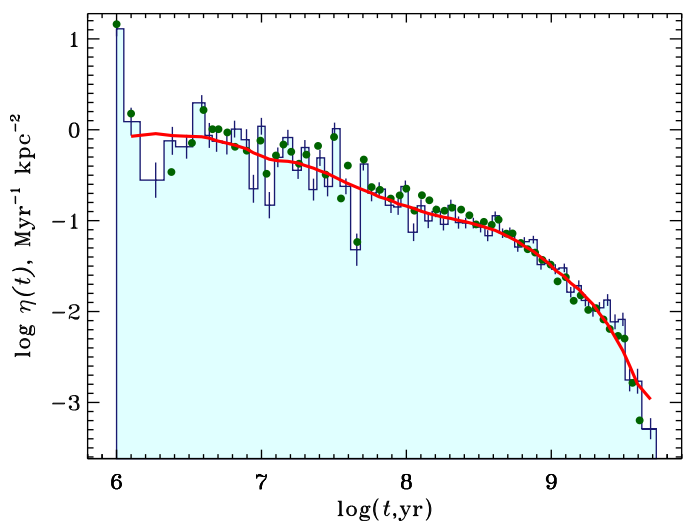}
\caption{Comparison of age distributions computed with different approaches to the completeness distance calculation. The distribution computed with individual completeness distances (Eq.~\ref{eq:agedv}) is shown with a histogram, while that computed with a single-completeness distance common to all clusters of the survey (Eq.~\ref{eq:agedc}) is shown with green filled circles. The vertical bars show the statistical uncertainty (Poisson errors) of the bins. The red curve illustrates a smoothed histogram.   
}
\label{fig:den_age}
\end{figure}
%------------------------------------------~\ref{fig:his_nage}-------------------------------

\subsection{Age distribution construction}\label{sec:method}

We define the cluster age distribution $\eta(t)$ as a surface density of objects in the unit interval of age $t$:
\begin{equation*}
\eta(t)= \frac{1}{S(t)}\,\frac{\mrm{d} N(t)}{\mrm{d} t}\,,   
\end{equation*} 
where $\mrm{d} N(t)$ is number of clusters with ages between $t$ and $t+\mrm{d} t$ residing within the completeness area $S(t)$. It is related to the more convenient logarithmic age distribution 
\begin{equation}
 \nu(t) = \frac{1}{S(t)}\,\frac{\mrm{d}N(t)}{\mrm{d}\log t}  \label{eq:agedistr}
\end{equation}
via
\begin{equation}
 \eta(t) = \frac{\log e}{t}\,\nu(t).  \label{eq:etanu}
\end{equation}

If one adopts a single completeness limit $\hat{d}_{xy,0}=1.8$ kpc, valid for clusters of all ages (horizontal line in Fig.~\ref{fig:dist_age}), then $S(t) \equiv S_0=\pi\,\hat{d}_{xy,0}^2$, and Equation (\ref{eq:agedistr}) re-written in the discrete form simply reflects the distribution of cluster numbers $\Delta_k N$ within the completeness area (i.e. below the horizontal line):
\begin{equation}
\nu_k= \frac{1}{S_0}\,\frac{\Delta_k N}{\Delta_k \log t}\,,   \label{eq:agedc}
\end{equation}
where the age step $\Delta_k\log t$ can be a variable.
 
In the case of the variable completeness limit, the cluster density will be computed as the sum of partial densities $\varsigma=1/(\pi\,\hat{d}_{xy}^2)$ of clusters located within their proper completeness limits given by Eq.~(\ref{eq:dciksrel}), i.e.\ those with $d_{xy}\leqslant \hat{d}_{xy}$ (black dots in Fig.~\ref{fig:dist_age}):
\begin{equation}
\nu_k= \frac{1}{\Delta_k \log t}\sum_{i=1}^{\Delta_k N}\varsigma_i = \frac{1}{\pi\Delta_k \log t}\,\sum_{i=1}^{\Delta_k N}
\frac{1}{\hat{d}_{xy,i}^2}\,. \label{eq:agedv}
\end{equation}
Here we sum over $\Delta_k N$ black dots within the given age interval $\Delta_k \log t$. Note that in the case of the constant completeness limit $(\hat{d}_{xy,i}\equiv \hat{d}_{xy,0})$ Eq.~(\ref{eq:agedv}) is naturally reduced to Eq.~(\ref{eq:agedc}). 

\begin{table}[t]
\caption{Comparison of the present open cluster sample with literature data used for cluster age distribution construction.}
 \label{tab:age_sam}
\begin{tabular}{llcrl}
\hline\hline
\noalign{\smallskip}
No&Sample      & $\hat{d}_{xy}$ & $N$ & Basic source         \\
  &            & kpc          &     &          \\
\hline
\noalign{\smallskip}
1 &Wi71        & 1.0            & 70  & BF71           \\
2 &PM86        & 1.0            & 116 & Lund3          \\
3 &BC91\tfm{a} & 2.0            & 94  & Lund5          \\
4 &La05\tfm{b} & 0.6            & 114 & COCD           \\
5 &Pi06\tfm{c} & 0.85           & 259 & COCD           \\
6 &Mo13\tfm{d} & 3.0\tfm{e}     & 143 & ATLASGAL\tfm{f}\\
7 &Present     & 1.0$-$6.0      &2242 & MWSC    \\
\hline
\end{tabular}
\tablebib{
(Wi71) \citet{wiel71}; (PM86) \citet{pama86}; (BC91) \citet{bcd91}; (La05) \citet{lamea}; (Pi06) \citet{clupop}; (Mo13) \citet{mora13};\\
(BF71) \citet{beckfen71}; (Lund3) \citet{lynga3}; (Lund5) \citet{lynga5}; (COCD) \citet{clucat}; (ATLASGAL) \citet{mora13}.
}
\tablefoot{\tft{a}{Subset of bright clusters $I(M_V)<-4.5$;} \tft{b}{Subset of known clusters;} \tft{c}{\textit{cmp} subset;} \tft{d}{Inner clusters with $|l|\lid60\degr,\,|b|\lid1.5\degr$;} \tft{e}{Distance limit of the representative sample;} \tft{f}{Compilation of 17 lists on the basis of \citet{daml02} catalogue ver.3.1.} 
}
\end{table}

The resulting distributions computed with help of Eqs.~(\ref{eq:etanu}), (\ref{eq:agedc}) and (\ref{eq:agedv}) are shown in Fig.~\ref{fig:den_age}. One can see that despite a considerable difference in the numbers of used objects collected from different areas (1359 in the first case and 2242 in the second one), both distributions are very similar, and for most age bins only differ within the statistical uncertainty. The difference at the youngest ages is due to the binning effect enhanced by poor statistics within the completeness distance (see Fig.~\ref{fig:dist_age}). However, one can see a small bias at $\log t>9$, where counts along Eq.~(\ref{eq:agedv}) lead to a slightly more enhanced (by factor of order 1.5 at maximum) cluster density, which is outside the statistical uncertainty. We consider this a consequence of taking into account ``far'' old clusters located beyond the common completeness limit  $\hat{d}_{xy,0}$. Due to this effect, and also due to the better representation of remote clusters, which might be important for the study of spatial variations of the age distribution in the wider solar neighbourhood, we will use the approach of a variable completeness distance 
in the remaining part of the paper. 

\subsection{Comparison with the literature}\label{sec:cmplit} 

Formerly ages of Galactic star clusters were accumulated from individual efforts of workers analysing their CMDs. From time to time these non-homogeneous data were reduced to a unique scale of ages and catalogued into compiled lists. Thus an appearance of a new collection of cluster ages was usually accompanied by a follow up study of galactic cluster age distribution. Nowadays this is regularly modulated by the appearance of large scale photometric and/or kinematic data which force generation of new sets of cluster ages.

%----------------------------------------------------------------------------%
\begin{figure}[t]
   \centering
\includegraphics[width=0.99\hsize,clip=]{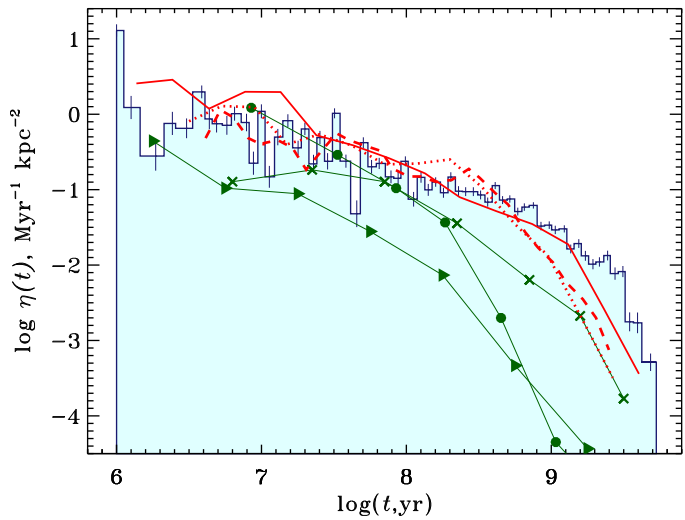}
\caption{Comparison of present (histogram) and literature age distributions. The vertical bars show the statistical uncertainty (Poisson errors). The six lines show age distribution samples from the literature presented in Table~\ref{tab:age_sam}. Different (green) symbols connected with thin lines show results from earlier published data. The filled circles, crosses, and triangles correspond to the Wi71, PM86, and BC91 age distributions, respectively. The thick (red) lines correspond to more recent cluster age distributions: the dashed line shows the La05 result, the dotted line is constructed from the Pi06 sample, and the solid line corresponds to the Mo13 data.   
}
\label{fig:cmp_litnt}
\end{figure}
%------------------------------------------~\ref{fig:his_nage}-------------------------------

\citet{wiel71} explored two catalogues of open cluster data \citep{beckfen71,lindoff68} and concluded that they give statistically similar age distributions. Here we consider the distribution based on the data of \citet{beckfen71}. The ages were based on the age calibration of \citet{barea69} and data on the colors and spectral classes of the brightest and bluest main sequence stars of the clusters. The clusters containing stars with spectral classes earlier than B2 were excluded from consideration. The analysis of the spatial distribution has shown that the samples are statistically complete within a cylinder with a radius of 1 kpc. The investigation of \citet{pama86} was based on the later Lund3 catalogue \citep{lynga3}. According to their conclusion, the sample is statistically complete within 1 kpc from the Sun, where it contains 116 objects (see Table~\ref{tab:age_sam}). The ages were also taken from Lund3. Clusters younger than 10 Myr were omitted. The fifth release of the Lund catalogue \citep{lynga5} was used for building the cluster age distribution by \citet{bcd91}, who used the subset of bright open clusters with integrated magnitudes $I(M_V)<-4.5$ mag. They found that this sub-sample could be regarded as spatially complete within 2 kpc of the Sun. 

In spite of the assurances on the completeness of the cluster samples mentioned above, more recent developments have shown that the real number of clusters in the solar neighbourhood is considerably higher than those presented by \citet{beckfen71} and the Lund collections. For example the COCD catalogue \citep{clucat,newc109}, based on the ASCC-2.5 \citep{asccnir} survey, when re-scaled to 1 kpc completeness distance, contains about three times more clusters than assumed by the previous studies. The MWSC survey with an average completeness limit of 1.8 kpc contains 400 objects within 1 kpc. Recently \citet{mora13} have considered an age distribution of star clusters from the inner Galactic disk. They have compiled a list of 695 known embedded and optical clusters located within the limits of the sub-millimetre survey ATLASGAL ($|l|\lid60\degr,\,|b|\lid1.5\degr$). They studied the completeness of the constructed sample and found that it is complete within 1 kpc from the Sun, and that it can be regarded to be representative within 3 kpc. 

%----------------------------------------------------------------------------%
\begin{figure}[t]
   \centering
\includegraphics[width=0.50\hsize,clip=]{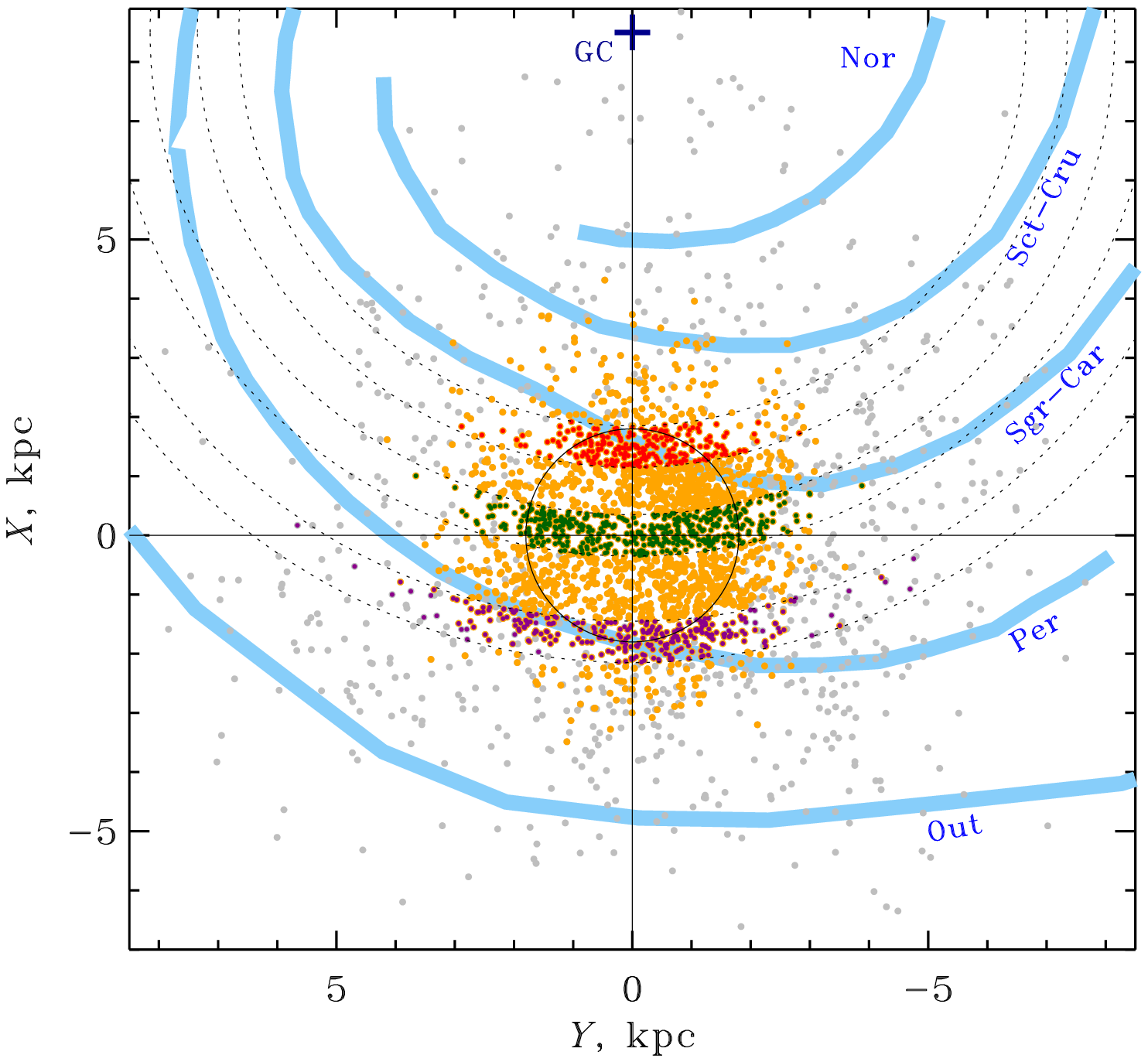}
\includegraphics[width=0.44\hsize,clip=]{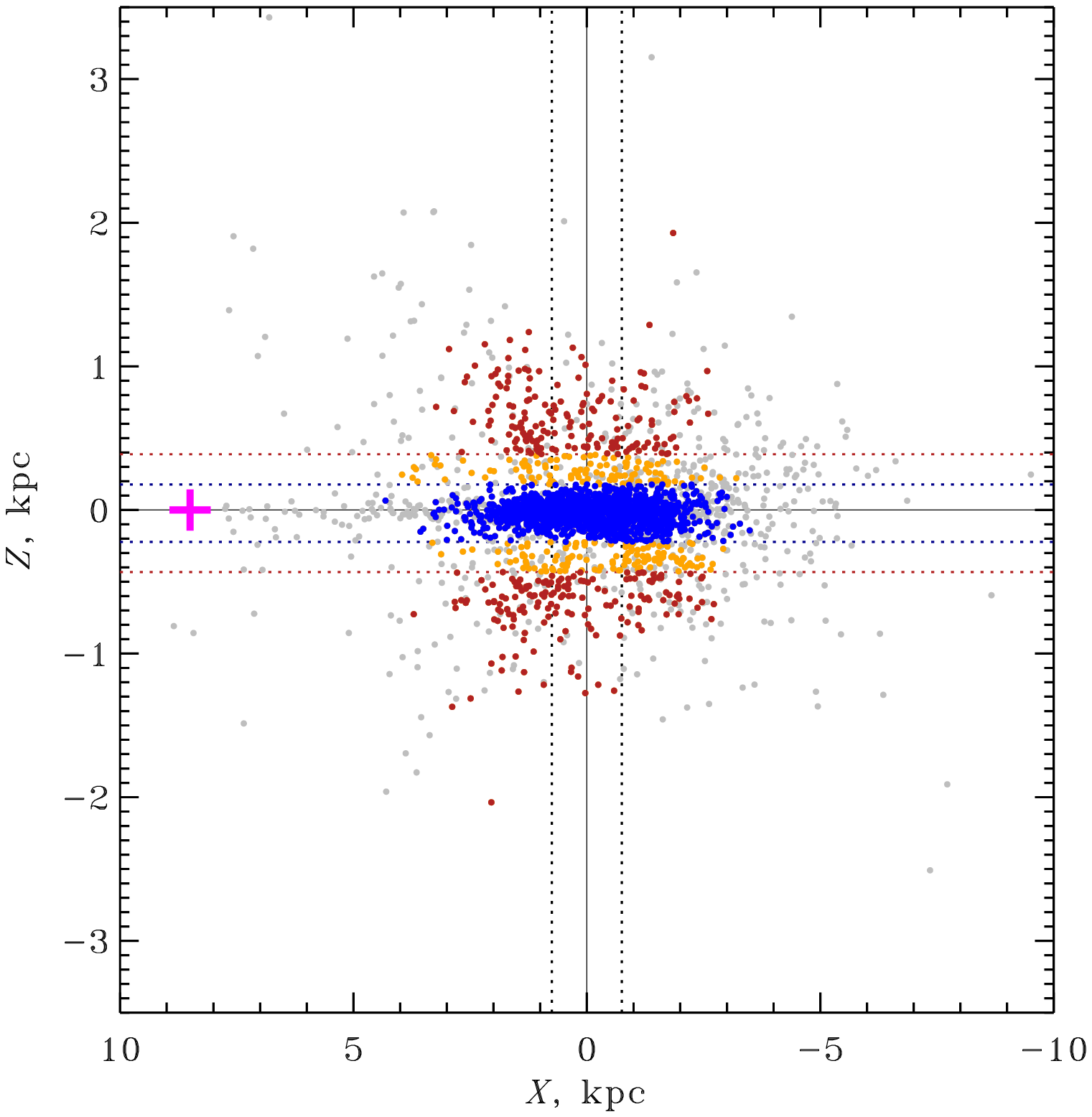}
\caption{Positions of the cluster spatial sub-samples considered. The left panel shows the ``planar'' samples and the right panel shows the ``vertical'' ones. Gray dots show all MWSC clusters, while coloured dots correspond to the completeness samples. Red, magenta, green, blue, and brown dots correspond to the Inner, Outer, Local, Thin-disk, and Thick-disk sub-samples, respectively. Their limits are shown with dotted lines. Yellow dots indicate the rest of the general completeness sample described in Sect.~\ref{sec:genage}. The big plus sign marks the Galactic centre. Thick curves show approximate positions of the spiral arms \citep[as taken from][]{benjam08}.
}
\label{fig:spat_smp}
\end{figure}
%------------------------------------------~\ref{fig:his_nage}-------------------------------

The aforementioned data samples are summarised in Table~\ref{tab:age_sam}, where we list the identifier of the sample (second column), the adopted completeness distance (third column), the reported number of clusters used for the age distribution construction (fourth column), and the basic source of open cluster data.

The comparison of the above distributions with present data is shown in Fig.\ref{fig:cmp_litnt}. The thick curves correspond to the determinations based on recent data. In general they show better agreement with the present determination than the thin curves corresponding to earlier publications, which demonstrates a general underestimation of cluster density. We attribute this to the insufficient completeness of these samples and to the already mentioned additional selection constraints imposed on the samples. The recent samples show general agreement with the present distribution for ages younger than $\log t<8.7$, and increasing deficiency at older ages. We attribute this bias to the NIR nature of the MWSC, which allows a better representation of old clusters containing bright red giants (see Sect.~\ref{sec:datcompl} for details). This might also explain the better agreement between our data and those of \citet{mora13}, also based on a survey including infrared data. We note an excess in the Mo13 distribution at young ($\log t<7.3$) ages. Among other reasons this could be caused by an enhanced cluster formation rate in the recent past in the inner galactic disk. Unfortunately, \citet{mora13} did not provide details of the construction of their age distribution, so we cannot discuss this feature as a possible consequence of their data analysis. Therefore, we postpone the discussion until Sect.~\ref{sec:spvar}, where we consider the issue of spatial variation of cluster age distributions of the MWSC clusters.

%----------------------------------------------------------------------------%
\begin{figure*}[t]
\includegraphics[width=0.69\hsize,clip=]{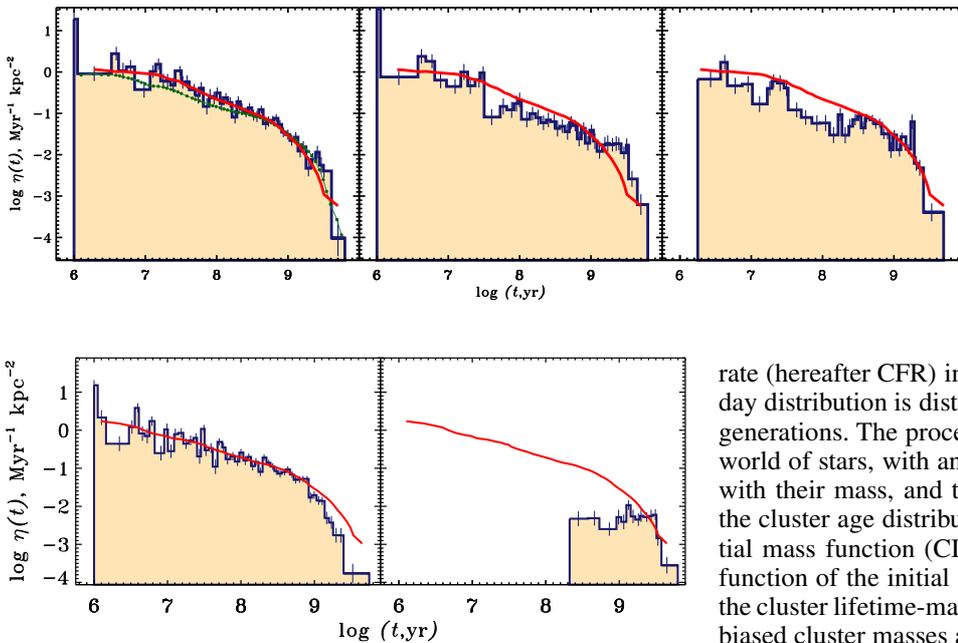}
\parbox[b]{0.30\hsize}{
\caption{Comparison of age distributions of various radial samples. The left, middle, and right panels show the Local, Inner, and Outer sample distributions, respectively. The thick red line is a smoothed Local distribution. It is plotted in the middle and right panels for comparison. The green curve in the left panel is a smoothed age distribution for the entire complete sample shown in Fig.~\ref{fig:den_age}.}\label{fig:dnrgvar}
}
\end{figure*}
%-------------------------------------------------------------------------

Our general conclusion from comparison with the literature is that the agreement of $\nu(t)$ with recent data is satisfactory and disagreement in the details is understandable. We also interpret the poor agreement with earlier results as a consequence of stronger incompleteness in the input catalogues burdened by additional constraints.

\section{Spatial variations of cluster age distribution}\label{sec:spvar} 

From Fig.~\ref{fig:dist_age} it is clear that depending on cluster age the MWSC sample is complete at solar-centric distances from 2 to 4 kpc. This allows us to trace variations in the cluster age distribution in a wide range of Galactocentric distances from about 6 to 12 kpc (which covers a significant part of the Galactic disk radius), and over the complete extent of the disk in the direction perpendicular to its plane. Taking into account the dependence of the completeness distance on age we have to note, that the complete age range can only be covered for the local Solar Neighbourhood closer than 2 kpc. The full span of the aforementioned distances is only available for clusters older than $\log t \approx 8.3$. This excludes the data on the radial dependence of the immediate cluster formation rate from our consideration, but still allows us to look for radial variations in the deeper history of cluster formation. 

\begin{table}[b]
 \caption{Spatial parameters of cluster samples}\label{tab:ssprm}
\begin{tabular}{@{}r@{ }lc@{ }c@{ }r@{}rc@{}r}
\hline\hline
\noalign{\smallskip}
 &Sample            &  $p$ &   $q$   &\mc{2}{c}{Range\tfm{a}}& Mean & $N_{obj}$\\
 &                  &      &         &         &      & position\tfm{a}&   \\
 &                  &  kpc &kpc/mag  &\mc{2}{c}{kpc}  & kpc         &      \\
\hline
\noalign{\smallskip}
1&Complete          & 0.80 & $-$0.42 & 4.2,   &12.0   & $\;\;$8.6   & 2242 \\
2&Inner             & 1.09 & $-$0.28 & 6.6,   &7.3    & $\;\;$7.1   & 254  \\
3&Local             & 0.80 & $-$0.42 & 8.2,   &8.9    & $\;\;$8.5   & 467  \\
4&Outer             & 0.56 & $-$0.57 & 10.0,  &10.7   & $\;$10.2    & 288  \\
\hline
\noalign{\smallskip}
5&Thin disk\tfm{b}  & 0.80 & $-$0.42 &$-$0.22,&0.18   & $-0.02$     & 750 \\
6&Thick disk\tfm{b} & 0.80 & $-$0.42 &$-2.04$,&$-0.43$& $-0.67$     &  95  \\
 &                  &      &         & 0.39,  &1.93   & $\;\;$0.65  &      \\
\hline
\end{tabular}
\tablefoot{\tft{a}{Samples 1-4 are in galactocentric radius, while samples 5,6 are in $Z$-coordinate.}
\tft{b}{Local sub-sample.}}
\end{table}

\subsection[]{Defining the spatial sub-samples}\label{sec:spasam}

In order to investigate the spatial stability of the age distribution in the Galactic disk we divided our completeness sample into a few spatially limited groups. Our division represents a compromise between the aim to reach maximum spatial separation of the groups, and to keep their population sufficient for reliable statistics. We have considered ``planar'' and ``vertical'' divisions. Following the geometrical considerations we selected our planar sub-samples in radial rings of given Galactocentric radii $R_G$. We include into the planar groups all the clusters independent of their distance from the Galactic plane. The ``vertical''  groups are separated with respect to their position along $Z$-axis by horizontal layers parallel to the Galactic disk plane. As a result, we have constructed five spatial sub-samples characterising the cluster population at areas spanning over $R_G\approx7$ to 11 kpc denoted here as Inner, Local, Outer, Thin-disk, and Thick-disk sub-samples. Their parameters are shown in Table~\ref{tab:ssprm}, where we also show data for our completeness sample discussed in Sect.\ref{sec:method} for comparison. In order to keep the general sampling approach we selected only local clusters for the vertical samples. To increase the statistics of Thick-disk clusters the width of the shell was increased to 1.5 kpc, as shown in the right panel of Fig.~\ref{fig:spat_smp}.   

We should note that an indication of non-isotropic behaviour in the completeness parameters $p$ and $q$ of Eq.~(\ref{eq:dciksrel}), which leads to variations of the completeness distance with galactic longitude, was already found in \citetalias{mwscint}. As the Inner and Outer sub-samples reside close to the border of the completeness zone, the issue of the completeness becomes especially important, so for these groups of clusters we decided to apply the values of $p$ and $q$ coefficients determined in \citetalias{mwscint} specifically for these directions. This is why the completeness distances do not coincide for the Inner and Outer samples, being shorter towards the Galactic centre, and longer in the opposite direction. For the other sub-samples we used the general values of $p$ and $q$ as shown in Table~\ref{tab:ssprm}. In Table~\ref{tab:ssprm} the radial or vertical limits of the groups, their average $R_G$ and $Z$-coordinate, and the number of included clusters are also provided.   

In Fig.~\ref{fig:spat_smp} we illustrate the spatial distribution of the constructed samples in the $(X,Y)$ and $(Z,X)$-planes. To give an impression of the covered portion of the Galactic disk we mark the position of the Galactic Centre and the approximate locations of the Galactic spiral arms given by \citet{benjam08}. As seen from the plot, the Inner and Outer samples roughly coincide with the positions of Sagittarius and Perseus spiral arms, while the Local sample represents the inter-arm cluster population.

\subsection[]{Planar variations}\label{sec:plavar}

In Fig.~\ref{fig:dnrgvar} we compare the age distributions of clusters from the three samples with different Galactocentric radii. The Local sample shows the $\eta(t)$ close to the general distribution that was discussed in the previous section. It covers nearly the same range of ages, with the exception of the oldest clusters in the last bin with $\log t \approx 9.7$. We attribute this to spatial sparseness of the oldest clusters, requiring a considerable extension of the area to collect a sufficient number of these objects. It can also be seen that the number of clusters younger than a few hundred million years significantly (by factor of 1.5-2) exceeds that of the general distribution. This is related to a bias due to incompleteness at the edge of the completeness zone and will be discussed in more detail below.

%----------------------------------------------------------------------------%
\begin{figure}[t]
\includegraphics[width=\hsize,clip=]{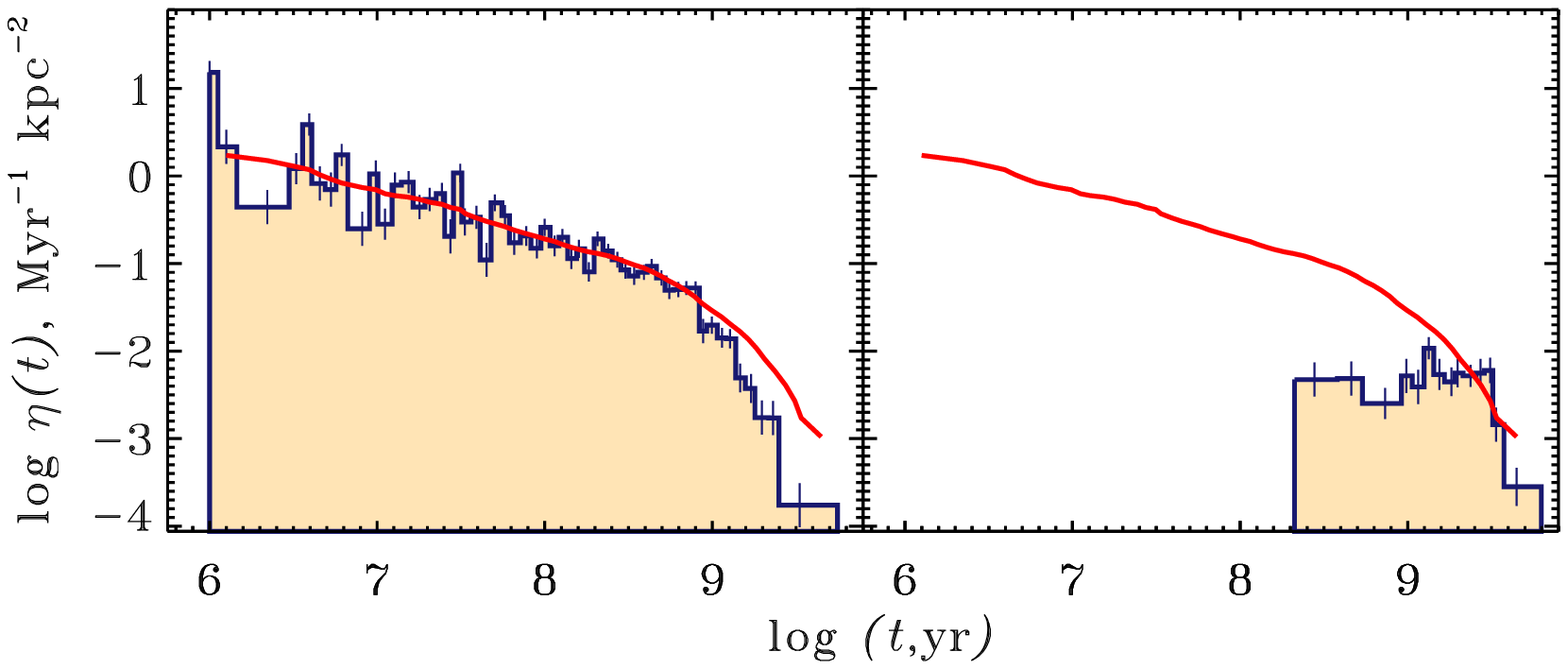}
\caption{Comparison of age distributions for thin-disk (left) and thick-disk (right) cluster populations shown with filled histograms. The thick red curve, as in Fig.~\ref{fig:dnrgvar}, shows the smoothed local distribution integrating all local shell clusters residing at different $Z$-coordinates.}\label{fig:dndtvert}
\end{figure}
%-------------------------------------------------------------------------

The Inner clusters show a similar distribution with only one difference from that of the Local sample. The Inner age distribution shows a deficiency at a moderately young cluster domain of $\log t\approx 7.7-8.7$. For younger ages, both distributions agree well, but at the older age domain $\log t\gtrsim 9$, the Inner distribution shows considerable excess with respect to the Local one. The Outer distribution also roughly resembles the basic features of the local distribution, and similar to the Inner sample exhibits a dip for the moderate ages. However, unlike the Inner clusters, an enhancement at older ages is not observed. Lastly, in contrast to both the inner and local distributions, the outer one shows a deficiency in young clusters with $\log t\approx7$ complemented with a total absence of the youngest clusters ($\log t<6.2$), which are abundantly present in the Local and Inner samples. 

All three samples show general agreement of the distributions representing both the inter-arm space and two different spiral arms. The different details observed in the distributions may reflect both different cluster formation histories and sampling biases. However, the latter is unlikely to be associated with an excess of older clusters in the inner sample, but rather with a higher cluster formation rate in the past in the inner disk. The deficiency of the youngest clusters in the outer disk may be due to a lower present cluster formation activity in the Perseus arm, and/or to the more difficult observing conditions of younger clusters behind heavy nearby clouds in the Perseus-Taurus-Auriga region. At the same time, it seems that the cluster formation histories of the Local and Outer samples were similar. We interpret the intermediate age dip as evidence of the increasing incompleteness among Pleiades-type clusters (those lacking bright stars, and especially red giants) at the edges of the completeness zone as it is illustrated by Fig.~\ref{fig:dist_age}.  

\subsection[]{Vertical variations}\label{sec:vervar}

In Fig.~\ref{fig:dndtvert} we show the age distributions of clusters of the ``vertical'' samples. In contrast to the general similarity shown by the ``planar'' samples (see Fig.~\ref{fig:dnrgvar}), the ``vertical" samples demonstrate a dramatic, but unsurprising, disagreement. As expected, the thin-disk distribution agrees closely with the ``planar" samples, and the thick-disk distribution is completely deprived of young clusters ($\log t<8.4$), with intermediate-age objects representing only a small fraction of the thick-disk population. For example, the fraction of objects with $\log t< 9$ is less than 20\% for the thick-disk, meanwhile for the thin-disk it exceeds 90\%. Both distributions complement each other when reproducing the total age distribution of the disk clusters, and can be regarded as representatives of different populations having different formation histories.

\section{Cluster formation history}\label{sec:cluhis}

In this section we present a simple analytic cluster formation and destruction model in order to discuss the impact of the different input parameters on the observed age distribution. The age distribution of star clusters reflects directly their formation history only in the regime where their lifetimes $\tau$ exceed the look-back time. Since the observed age distribution is monotonically declining, a consequence would be an increasing cluster formation rate (hereafter CFR) in the recent past. Otherwise their present-day distribution is distorted by destruction processes of existing generations. The process is quite similar to that observed in the world of stars, with an exception, that stellar lifetimes decrease with their mass, and those of star clusters increase. In general the cluster age distribution depends on the CFR, the cluster initial mass function (CIMF) and the cluster lifetime, which is a function of the initial cluster mass and will be parametrised by the cluster lifetime-mass relation (LTMR). Since reliable and unbiased cluster masses are not available to date, the details of the gradual cluster dissolution do not directly enter the present day age distribution. The LTMR describes the observability of clusters, where we allow for a initial mass dependent maximum age of the clusters. For a proper interpretation of the observed age distributions these dependencies have to be taken into account. Our model has some free parameters, which should be optimised when the model is fit to the empirical age distributions.

\subsection{The model}\label{sec:model}

Let $\xi(M,T)\,\mrm{d}M\,\mrm{d}T$ be the number of clusters with initial masses $M,\,M+\mrm{d}M$ formed in the time interval $T,\,T+\mrm{d}T$
\begin{equation*}
\xi(M,T) \equiv \frac{\partial^2N}{\partial T\partial M}. 
\end{equation*}

The time $T$ is counted from the moment of formation of the open cluster subsystem of the Galactic disk that is still observable. To be definite we assume that this moment corresponds to formation of the oldest cluster in the completeness sample with $\log t_{max}=9.68$\footnote{In fact this limit corresponds to the value of the last bin in the age distribution and is slightly lower than the age of the oldest cluster.}. In this scale the present moment of time $T_p$ equals 4.8 Gyr. The cluster mass $M$ corresponds to the initial mass of the cluster, and normally decreases with cluster evolution due to mass loss driven by various processes. In the literature starting with the seminal works of \citet{salp55} and \citet{mschm59} $\xi(M,T)$ is called the ``cluster formation function'' and is typically
represented as a product of two independent functions of mass and time
\begin{equation*}
\xi(M,T)  = \psi(T)\,f(M)\,, 
\end{equation*}
where $\psi(T)$ is the CFR, and $f(M)$ is the CIMF. The CIMF is normalised to unity over the whole mass range $[M_{min},M_{max}]$
\begin{equation}
\int^{M_{max}}_{M_{min}} f(M)\, \mrm{d}M = 1\,,    \label{eq:norm}
\end{equation}
and while $\psi(T)$ gives the number of clusters formed per time interval, $f(M)$ weights it with cluster mass.  
 
%----------------------------------------------------------------------------%
\begin{figure}[t]
\begin{center}
\includegraphics[width=0.9\hsize,clip=]{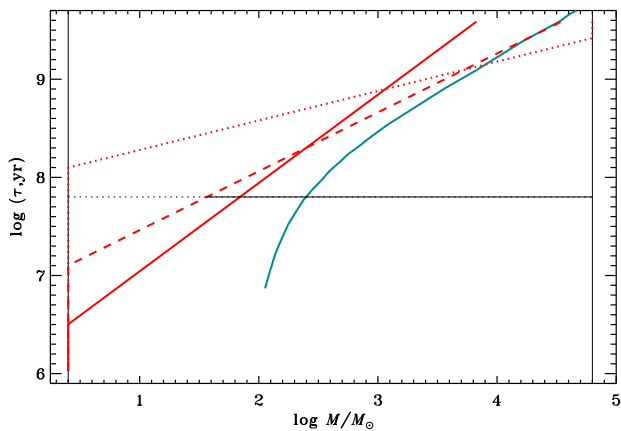}
\caption{The lifetime-mass relations LTMR used in the model (red) compared to the \citet{lamgi06} relation with 100 $M_\sun$-remnant (blue). The solid red line corresponds to under-filled clusters, the dashed line represents filling Roche lobe models, and the dotted line is for overfilled models with $T_0=500$ Myr. The two thin, vertical lines indicate lower and upper limits of the cluster mass range. The horizontal solid line at $\log \tau=7.8$ illustrates the integration range $[M_t,M_{max}]$ for the filled model.}\label{fig:mltrcmp}
\end{center}
\end{figure}

With these definitions and $\tau(M)$, the lifetime of a cluster with initial mass $M$ (the LTMR), one can easily build a theoretical distribution of cluster ages in terms of the cluster formation history. We take into account that the relation between the moment of cluster formation $T$ and its current age $t$ is $T=T_p-t$. The number $\mrm{d}N$ of clusters with mass $M,\,M+\mrm{d}M$ formed at a moment $T,\,T+\mrm{d}T$ and not dissolved until the present is equal to $\xi(M,T)\,\mrm{d}M\,\mrm{d}T$ if $\tau(M) \geqslant t$, and 0 otherwise. For simplicity we use hereafter the notation $M_t$, denoting the solution of the equation $\tau(M)=t$, corresponding to the minimum mass of presently observed clusters with age $t$. 

For clusters of all masses born at $T,T+\mrm{d}T$ the observed number is expressed as
\begin{equation*}
\mrm{d}N = \int^{M_{max}}_{M_{min}}\xi(M,T)\,\mrm{d}M\,\mrm{d}T - \int^{M_t}_{M_{min}}\xi(M,T)\,\mrm{d}M\,\mrm{d}T, 
\end{equation*}
where the second integral corresponds to the number of dissolved clusters at the moment $T_p$. Changing to cluster ages (= look-back time), the present-day age distribution becomes
\begin{eqnarray}
 \eta(t)=\frac{\mrm{d}N(t)}{\mrm{d}t} &=& \int^{M_{max}}_{M_t} \xi(M,T_p-t)\, \mrm{d}M\,  = \nonumber \\   \label{eq:modeta}
         &=& \psi(T_p-t)\,\int^{M_{max}}_{M_t} f(M)\, \mrm{d}M.      
\end{eqnarray}
As the age $t$ increases, the lower integration limit also increases from $M_{min}$ to $M_{max}$, while the integral decreases from unity to zero. At young ages it is clear that $\eta(t)$ is close to the CFR, but for old clusters it is highly affected by the assumptions on $f(M)$ and $\tau(M)$. 

\begin{table}[t]
 \caption{Fitted models and best fit parameters for the CIMF}
 \label{tab:fitcimf}
\begin{tabular}{llllllll}
\hline\hline
\noalign{\smallskip}
Model  &$N_{it}$&$N_{fd}$&$\chi^2_n$  &$x_1$ &$\sigma_{x_1}$&$x_2$&$\sigma_{x_2}$\\
%       Nit       Nfg      \chi        x1     sx1    x2     sx2   
\hline
\noalign{\smallskip}
$u$       & 10     & 36     & 1.343     & 0.39 & 0.18 & 0.54 & 0.05 \\
$f$       & 10     & 36     & 1.403     & 0.63 & 0.15 & 0.24 & 0.05 \\
$o$\tfm{a}& 177    & 33     & 1.524     & -    & -    & 0.07 & 0.05 \\
\hline
\end{tabular}
\tablefoot{\tft{a}{One section CIMF.}} 
\end{table}

Equation (\ref{eq:modeta}) fully determines the cluster population model describing the theoretical age distribution. The main components of the model are the cluster initial mass function, the cluster formation rate, and the cluster lifetime, along with their parameters. One can see from Eq.~(\ref{eq:modeta}) that the theoretical $\eta(t)$ depends on the specific representations of $f(M)$, $\psi(T)$ and $\tau(M)$, and on the parameters of these functions. In earlier studies, different approaches were proposed, depending on the purpose of the study and available data for the components. Below we briefly describe the adopted representation for every component. The theoretical age distribution was fit to the empirical one with the help of the powerful and flexible routine MPFIT from the IDL-library of \citet{markwdt09}. 

For the cluster formation rate CFR we use an exponential function 
\begin{equation}
 \psi(T) = \alpha + \beta\exp\left(\gamma \frac{T_p-T}{T_p}\right). \label{eq:cfr}
\end{equation}

Prior to selecting this particular form for the CFR, we tested various other forms (linear relation, rational, power law etc.) and found that the resulting goodness of fits do not differ strongly and there is no clear preference. Nevertheless in detail they differ and we decide to show here one giving a small specific residual to the fit $\chi^2_n$. In some regions of the parameter space $\gamma$ is strongly correlated with $\alpha$ and $\beta$, so the iteration does not converge properly. In these cases $\gamma$ is fixed to a few different values and the best fit results by optimization of the parameters $\alpha$ and $\beta$ are compared. Eq.~(\ref{eq:cfr}) reproduces a  CFR monotonically decreasing in time if $\beta$ and $\gamma$ are positive. At the initial moment we have $\psi(0)\equiv\psi_0=\alpha+\beta\,\mathrm{e}^\gamma$, while the present-time CFR is equal to $\psi(T_p)\equiv\psi_p=\alpha+\beta$. The average CFR $\psi_a$ can then be expressed as 
\[
\psi_a = \frac{1}{T_p}\int_0^{T_P} \psi(T)\, \mrm{d}T = \alpha + \frac{\beta}{\gamma}\,(\mrm{e}^\gamma-1).
\]
As a measure of the variations of the CFR we use two ratios: $\psi_0/\psi_p$ and $\psi_a/\psi_p$.

%----------------------------------------------------------------------------%
\begin{figure*}[t]
\includegraphics[width=0.325\hsize,clip=]{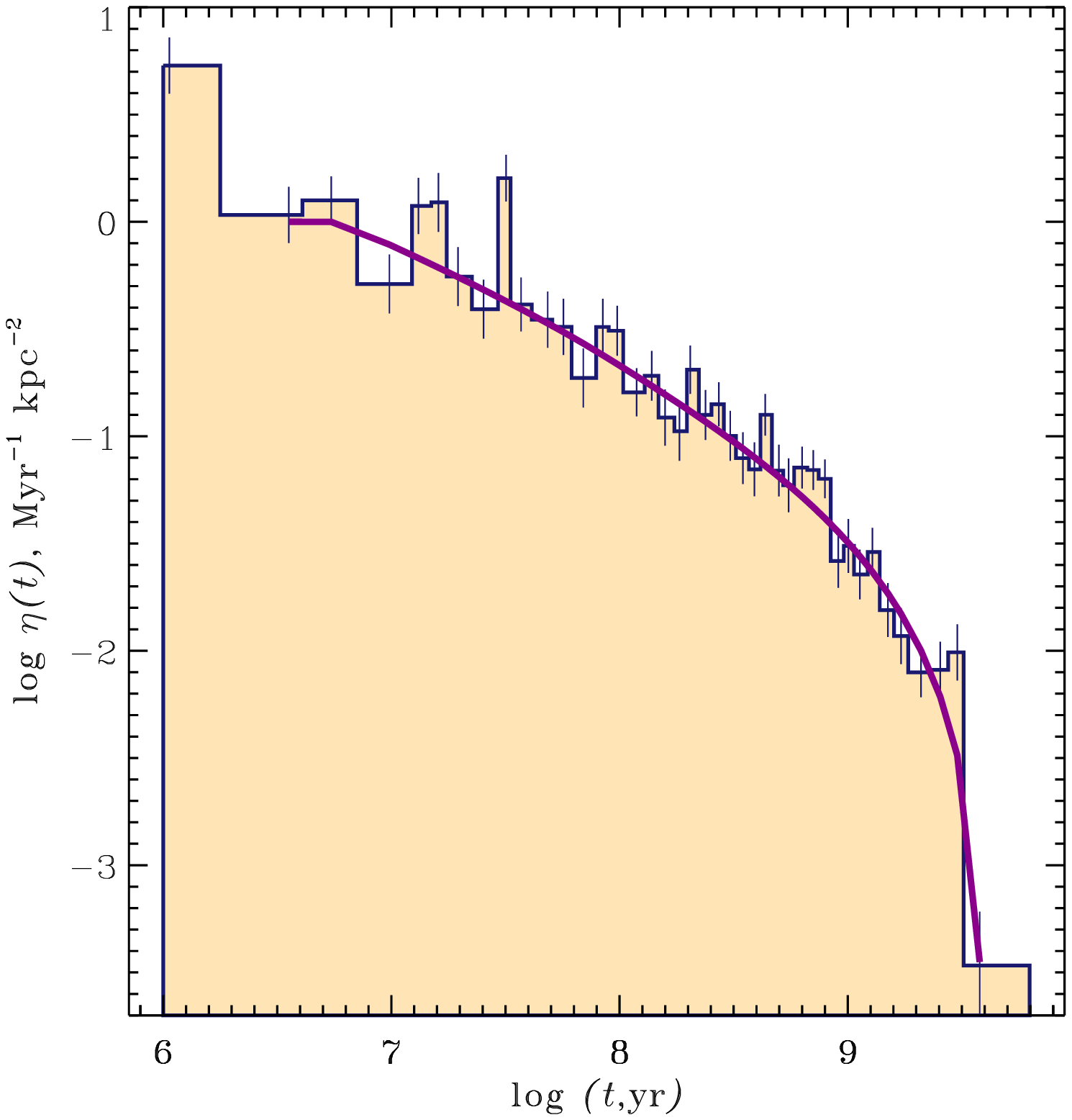}
\includegraphics[width=0.325\hsize,clip=]{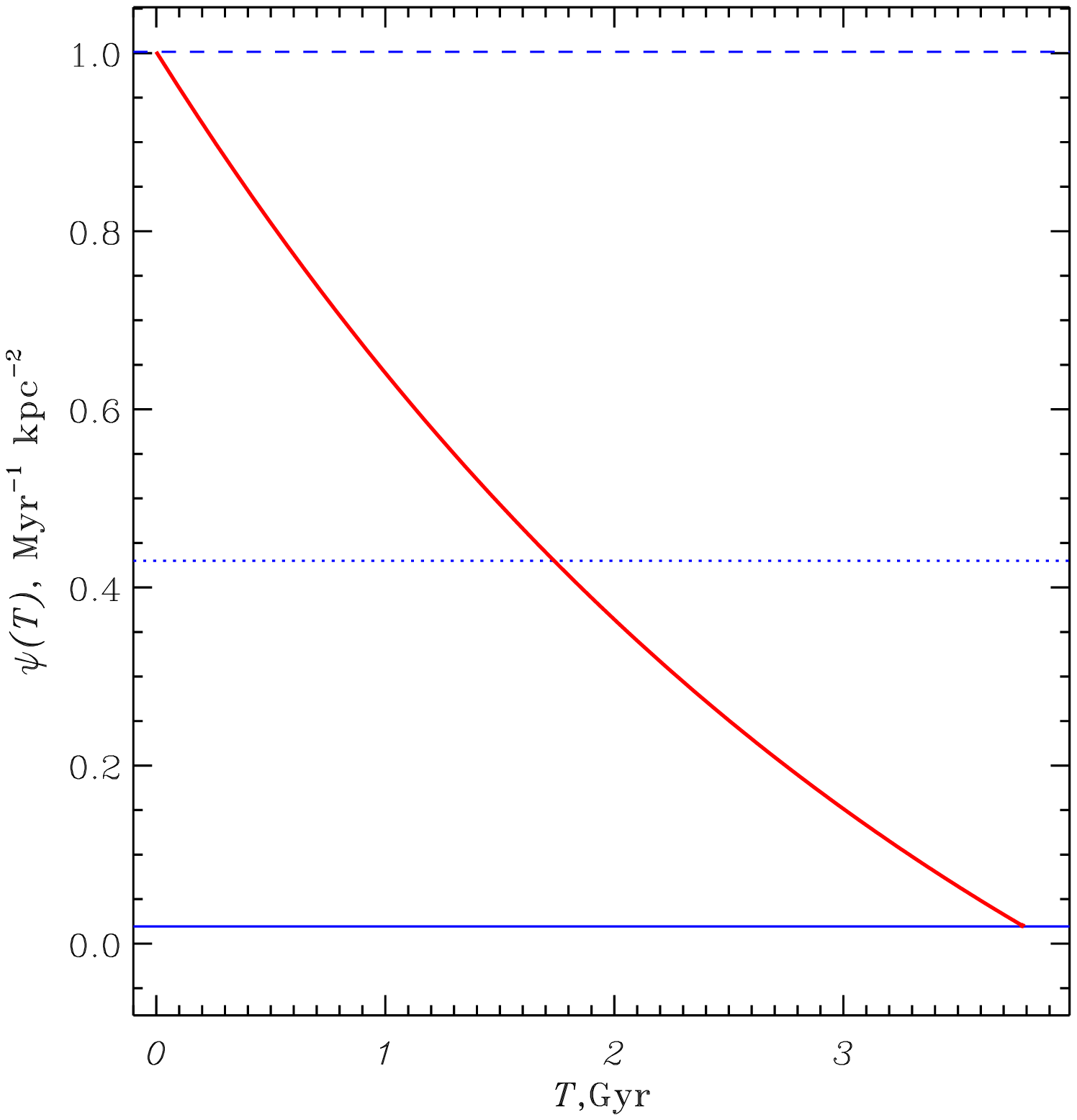}
\includegraphics[width=0.325\hsize,clip=]{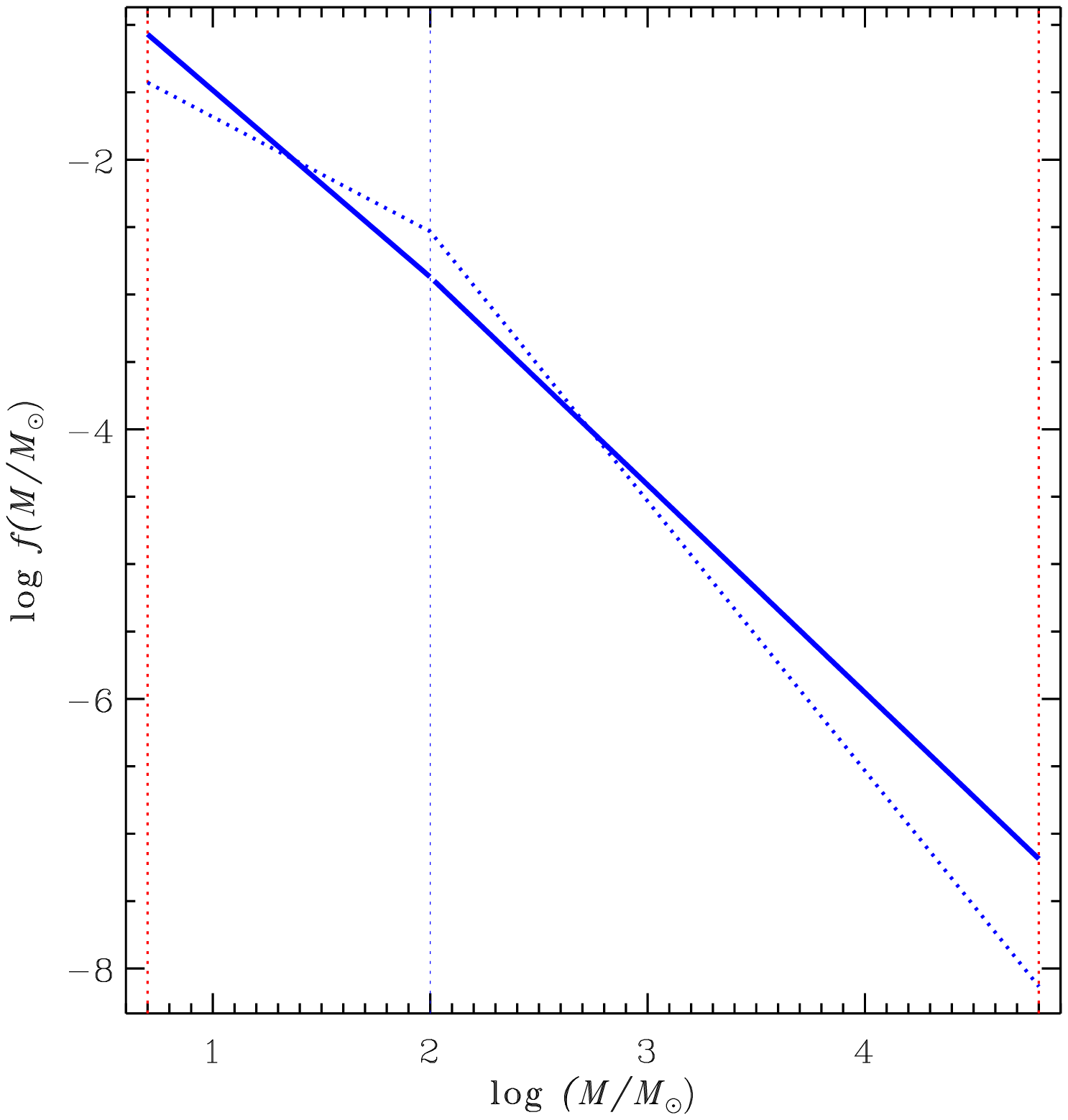}\\
\includegraphics[width=0.325\hsize,clip=]{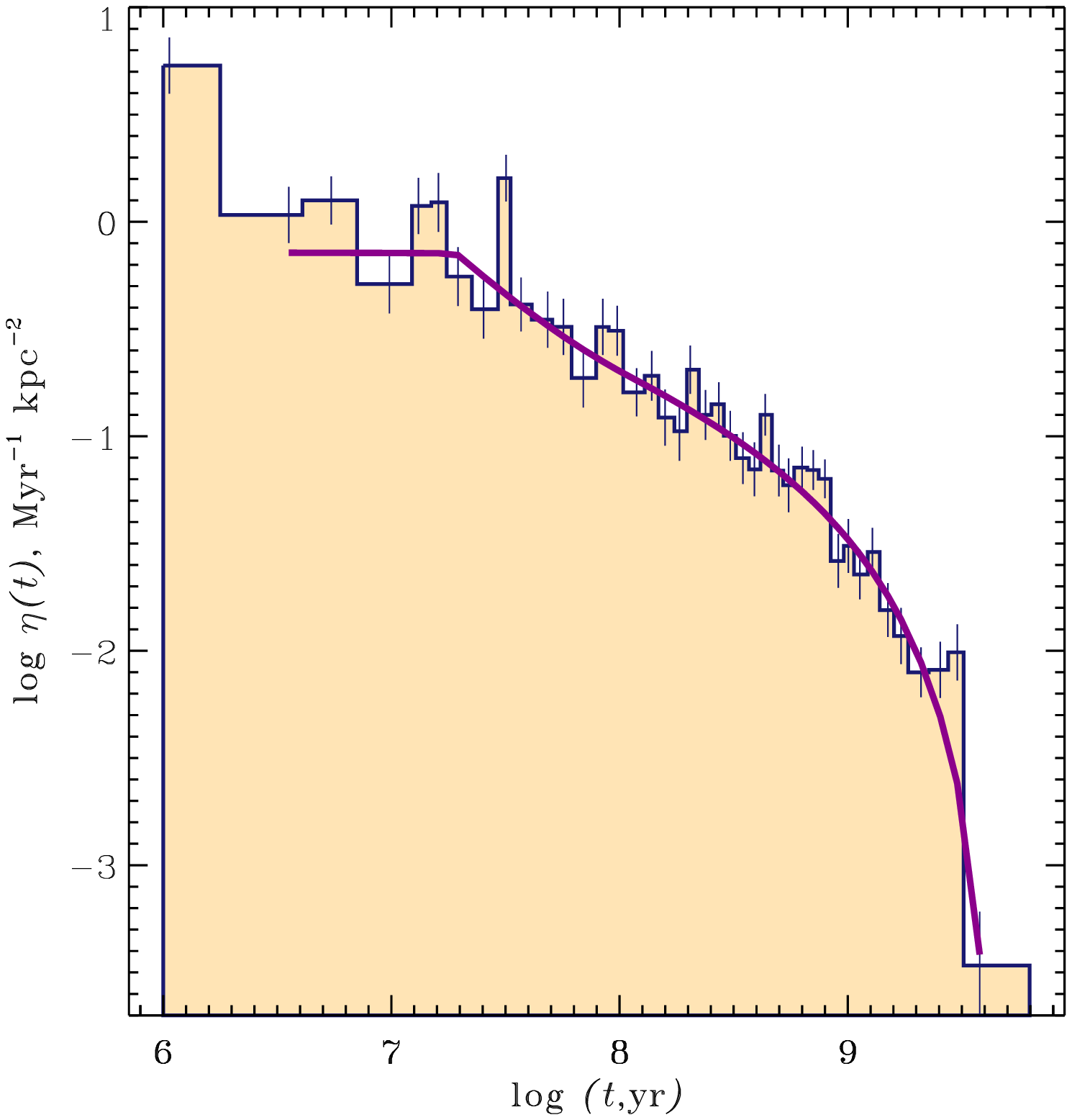}
\includegraphics[width=0.325\hsize,clip=]{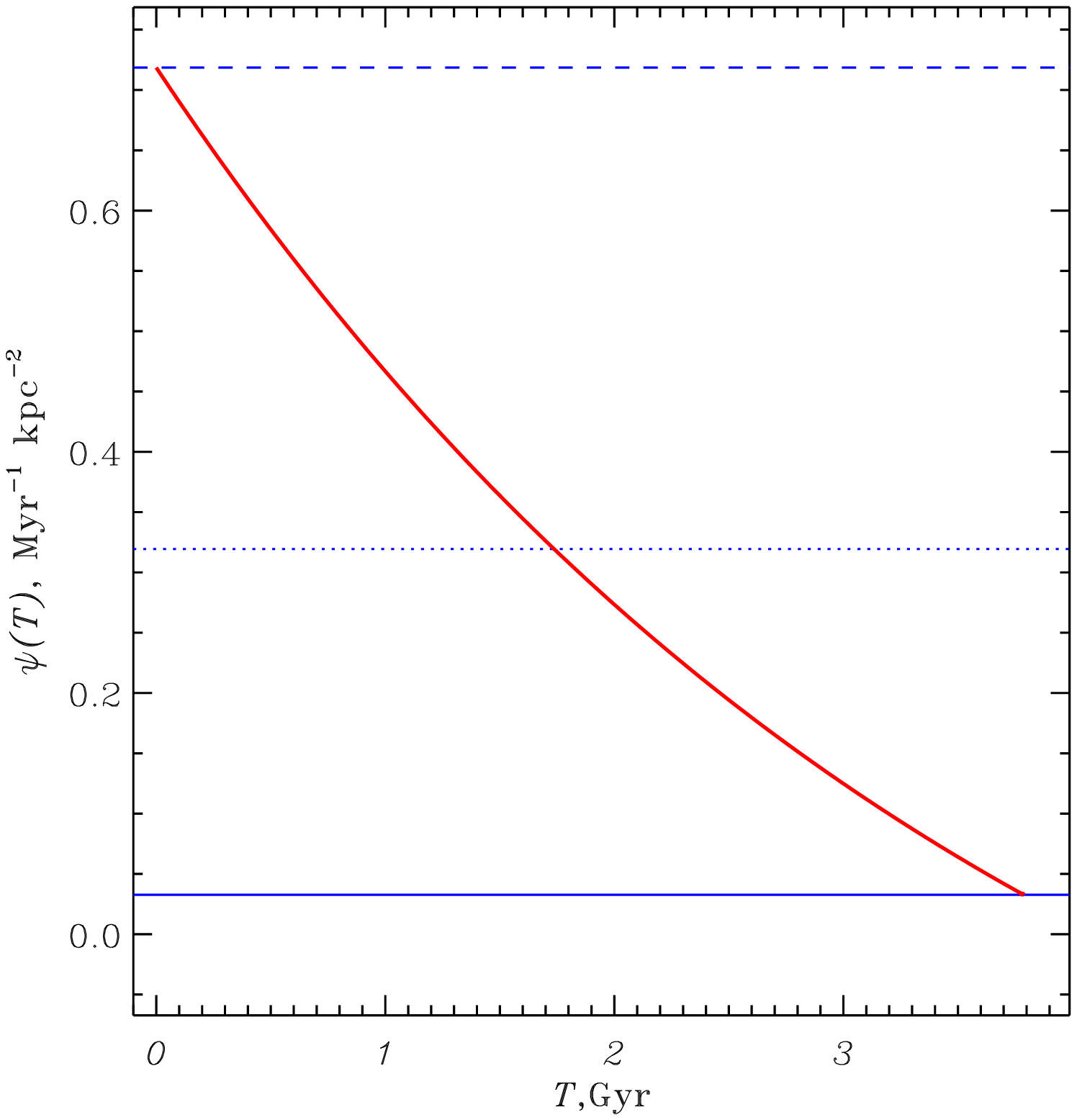}
\includegraphics[width=0.325\hsize,clip=]{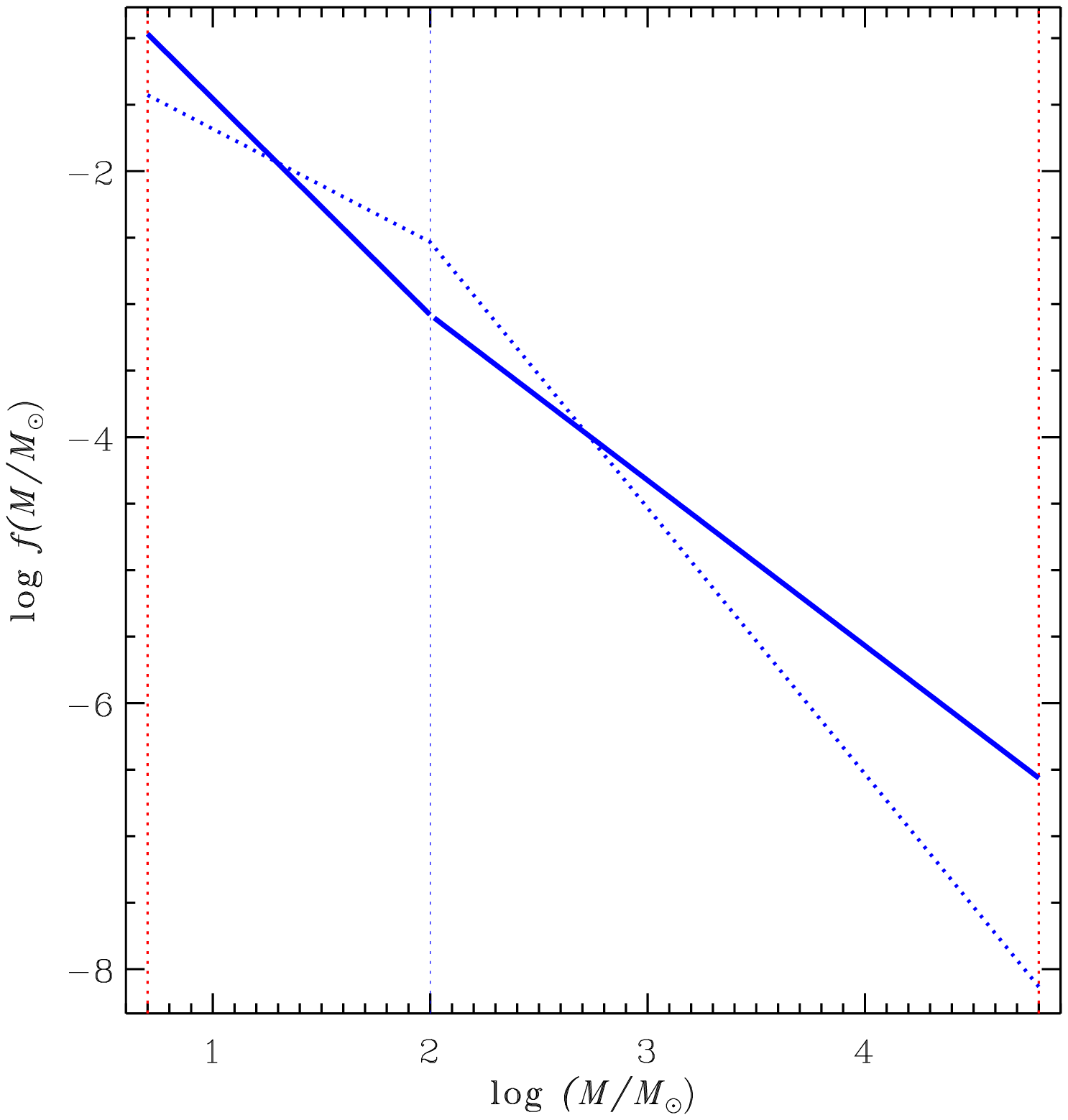}\\
\includegraphics[width=0.325\hsize,clip=]{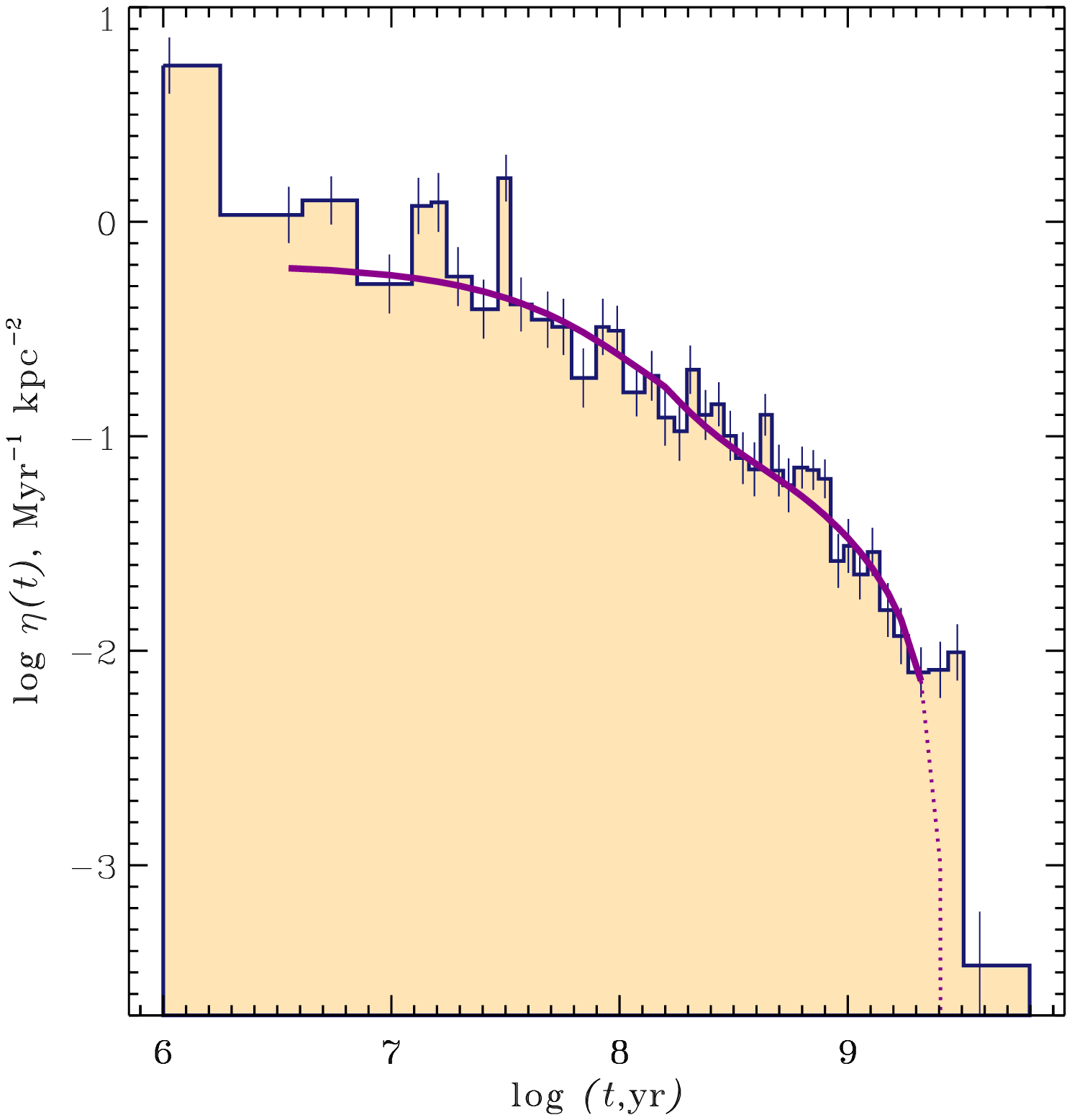}
\includegraphics[width=0.325\hsize,clip=]{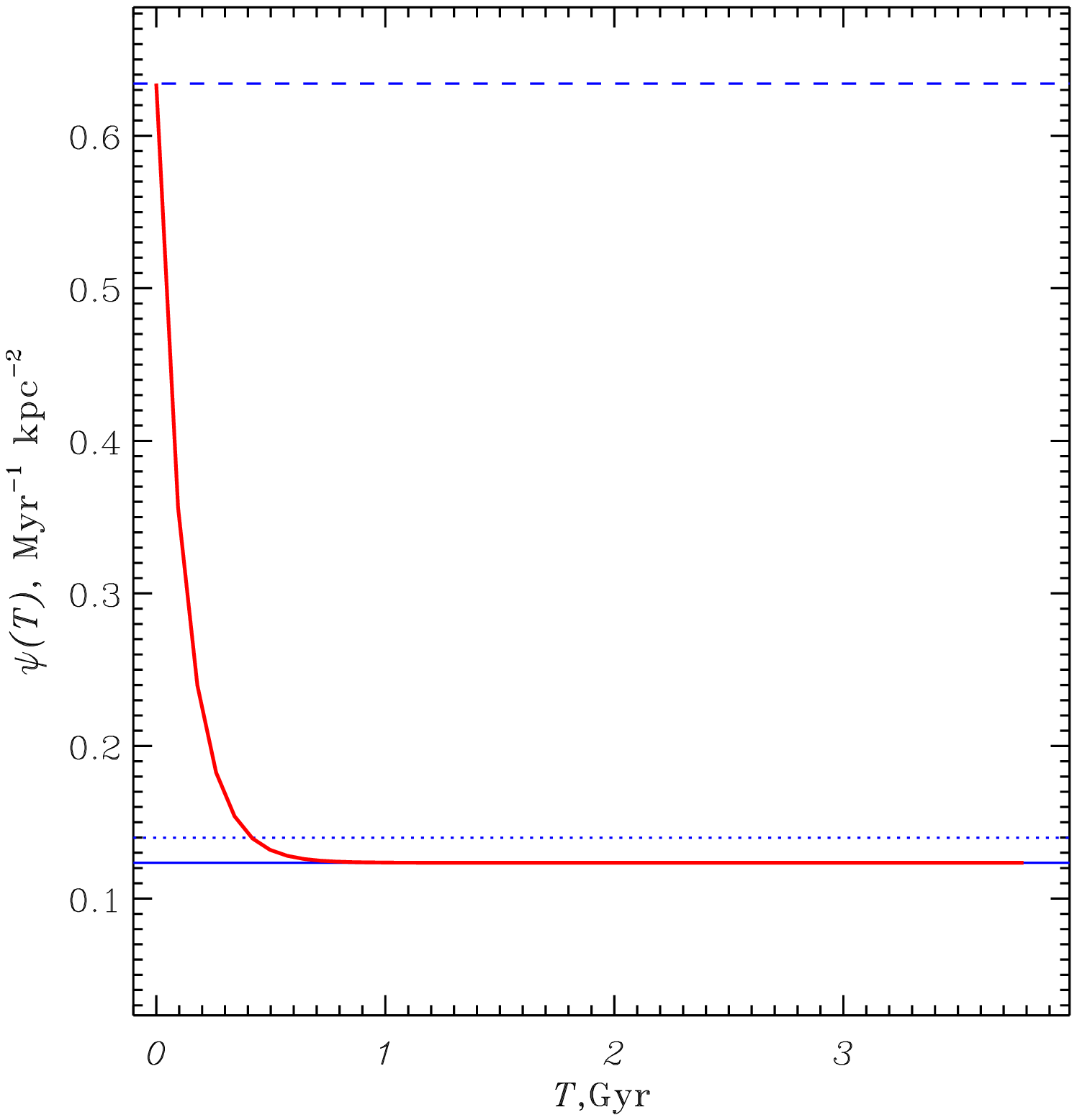}
\includegraphics[width=0.325\hsize,clip=]{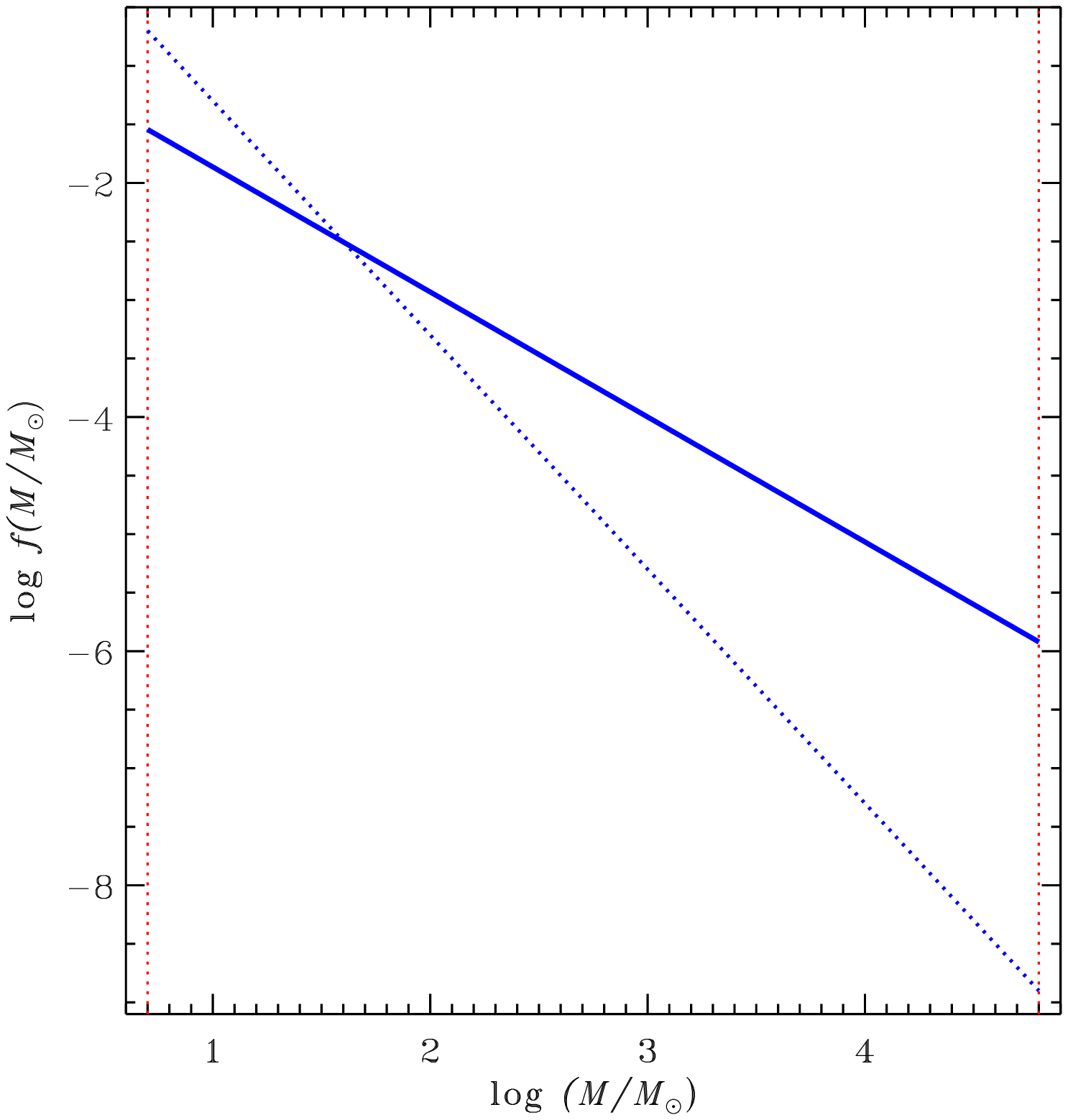}
\caption{Results of the model fit to the observed age distribution for the local cluster sample for $u$-, $f$-, and $o$-models (from top to bottom).  The left column shows the fit results, the middle column shows the respective CFRs, and the right column displays the derived CIMFs. The histogram shows the observed age distribution with Poisson errors indicated by vertical error bars. The violet curves represent the fitted models, where the thick portion indicates the fitted range, and the thinner dotted one extends the derived law. Red curves are the model CFRs, thick blue lines are the CIMFs and thick blue dotted lines represent their initial approximations. The blue horizontal lines indicate the present (solid), initial (dashed), and average (dotted) model CFRs. Vertical lines indicate the mass range limits (red), and selected $M^*$ (blue) value used.}\label{fig:ufores}
\end{figure*} 

In principle the CIMF agrees with the present day mass distribution of very young clusters, before they suffer from mass dissolution. A simple representation of the CIMF is given by a broken power law with two sections

\begin{eqnarray}
 f(M) = \frac{\mrm{d}N}{\mrm{d}M} = \begin{cases}
                                      k_1\,M^{-(x_1+1)} & \text{for $M_{min}\leqslant M < M^*$,}\\        \label{eq:cimf2}
                                      k_2\,M^{-(x_2+1)} & \text{for $M^* \leqslant M \leqslant M_{max}$.}
                                    \end{cases}  
\end{eqnarray}
The constants $k_1,k_2$ are determined by the continuity (or the jump) at $M^*$ and the normalisation of the CIMF with Eq.~(\ref{eq:norm}). The parameters $x_1$, $x_2$ are determined from the model fit to the observed age distributions, with initial values $x_1=-0.15$, and $x_2=1.0$. The mass ranges are fixed at the following values: $M_{min}=2.5\,M_\sun$, $M_{max}=6.3\,\times\,10^4\,M_\sun$ and $M^*=100\,M_\sun$. 

% qqfig

\begin{table*}[t]
 \caption{Best fit parameters for the CFR}\label{tab:fitcfr}
\begin{tabular}{lllllllllllllll}
\hline\hline
\noalign{\smallskip}
Model&$\alpha$&$\sigma_\alpha$ &$\beta$   &$\sigma_\beta$ &$\gamma$ &$\sigma_\gamma$  &$\psi_0$&$\sigma_{\psi_0}$&$\psi_a$&$\sigma_{\psi_a}$&$\psi_p$&$\sigma_{\psi_p}$&$\psi_a/\psi_p$ &$\psi_0/\psi_p$\\
%       alp       ealp       bet          ebet           gam       egam   psi0      epsi0      psia    epsia     psip        epsip    a/p    0/p
\hline
\noalign{\smallskip}
$u$   & $-0.55$& $0.10$ &  $0.57$       & $0.10$       &1.00\tfm{b}& -  &  $1.00$  &  $0.29$  & $0.43$ & $0.20$ &  $0.02$  &  $0.14$ & $22.2$& $51.7$\\
$f$   & $-0.37$& $0.05$ &  $0.40$       & $0.05$       &1.00\tfm{b}& -  &  $0.72$  &  $0.15$  & $0.32$ & $0.10$ &  $0.03$  &  $0.08$ & $9.8$ & $22.0$\\
$o$   & $0.12$ & $0.02$ &  $1.38$\tfm{a}& $0.22$\tfm{a}&31.24      &0.00&  $0.63$  &  $0.08$  & $0.14$ & $0.02$ &  $0.12$  &  $0.02$ & $1.1$ & $5.1$ \\
\hline
\end{tabular}
\tablefoot{\tft{a}{In units of $10^{-14}$; }\tft{b}{Fixed.}} 
\end{table*}

The third input function is the lifetime-mass relation LTMR. It is well-known from numerical simulations that the LTMR strongly depends on the initial conditions of star clusters after gas-removal. Here we use a simple parametrisation covering the results based on N-body calculations of \citet{ernstea15}, who studied the dissolution of star clusters of different initial Roche volume filling factors covering a large cluster mass range. 
They consider cases of under-filled, filled in, and overfilled Roche lobes. They have shown that the lifetime scales with a power of the relaxation time with a decreasing power law index for larger filling factors. We use here a simple power law of the initial mass
\begin{equation}
 \tau = T_0 \left(\frac{M}{M_0}\right)^{s},  \nonumber
\end{equation} 
with $s=0.9, 0.6, 0.3$ for the under-filling (u), filling (f) and overfilling (o) cases, respectively. The filling case is also very close to the parametrisation of  \citet{lamgi06}. The respective relations are shown in Fig.~\ref{fig:mltrcmp} together with earlier results of \citet{lamgi06}.

Since the issue of the Roche volume filling factor is not solved (in particular it is not clear how many clusters follow the extremely compact or the extended models), we will not use the parameters of the LTMR for the optimisation of the model. Instead, we simply compare the fit results for different filling cases with fixed $T_0=200$\,Myr for the under-filling and $T_0=500$\,Myr for the overfilling case at $M_0=250 M_\sun$. The larger $T_0$ for the overfilling case is necessary to reach a maximum age of a few Gyr for the most massive clusters.

\subsection{Results of the model fit}\label{sec:fitreslt}

The best fit parameters determined for the three cases of under-filling, filling and overfilling clusters are listed in Tables~\ref{tab:fitcimf} and \ref{tab:fitcfr}. The resulting fits are shown in Fig.~\ref{fig:ufores} (left column). For each model, the required number of iterations $N_{it}$, the number of freedom degrees $N_{fd}$, the specific (per one freedom degree) $\chi^2_n$-parameter describing goodness of fit, and the slopes of the CIMF with their errors are provided in Table~\ref{tab:fitcimf}. For overfilling clusters, a one-section representation of the CIMF is used with an initial slope $x=1.0$. In Table~\ref{tab:fitcfr} we list the CFR parameters $\alpha$, $\beta$, $\gamma$, and initial $\psi_0$, average $\psi_a$, and present-day $\psi_p$ values and their ratios. For the $u$- and $f$-models we use the full range of available ages for fitting. However, the $o$-model was fit at $\log t\lesssim 9.3$ since overfilling clusters with older ages should not exist for $M\leqslant M_{max}$ (see Fig.~\ref{fig:mltrcmp}).  

For the under-filling and filling cases we find a very good representation of the observed age distribution. In both cases we find a strong decrease in the CFR (see Fig.~\ref{fig:ufores} middle column) and a relatively shallow power law of the CIMF at high cluster masses (Fig.~\ref{fig:ufores} right panels). We have tested the fits with different values of $\gamma$ in the CFR and $T_0$ in the LTMR, but the results were all very similar. 
In contrast, the overfilling case does not yield a satisfactory fit with a 2-slope CIMF and fixed $\gamma=1$. With a 1-slope CIMF and free parameter $\gamma$ we find a reasonable fit (after increasing $T_0$ to 500\,Myr). The resulting CFR shows a strong initial peak on top of the dominating constant value over the majority of time. The main reason is that in this case the cluster lifetimes cover only a range of less than 1\,dex, which cannot reproduce the continuous decline of the observed age distribution over 2\,dex with a monotonically declining CFR. 

The deviation of the CIMF shape to the initial mass distribution is large in all three cases. It increases with a decreasing power law index $s$ of the LTMR, i.e. with an increasing filling factor of the clusters, leading to a very shallow function at the high mass end. 

\subsection{Discussion}\label{sec:discus}

The simple model for the three input functions, CIMF, CFR, and LTMR, described in the previous sub-section, yields a few fundamental conclusions for the observed present-day cluster sample.

\begin{itemize}
\item The age distribution alone cannot disentangle the impact of the three input functions. For a better understanding of cluster formation and evolution, a 2-dimensional fit of the mass-age distribution would be helpful. But this requires a parametrisation of the cluster mass evolution $M(t)$, replacing the simple function for the cluster lifetime.
\item The fits to the observed age distribution for the different Roche volume filling factors are indistinguishable. However, shallower LTMRs require a sharper peak in the CFR at the oldest ages.
\item In the framework of our simple model the CFR is not proportional to the field SFR as derived by \citet{aumebin09} or \citet{jujah10}.  
\item The large fraction of clusters with intermediate age combined with the strong decrease above an age of 1\,Gyr requires a large fraction of high mass clusters, i.e. a shallow CIMF at the high mass end, with $x_2$ significantly smaller than unity.
\item We do not find a significant break in the CIMF to be flatter at low masses. One reason could be the extrapolation to very short lifetimes at the low mass end. On the other hand, infant mortality or a significant mass loss due to the expulsion of gas after cluster formation is not taken into account here \citep[see e.g.][for violent relaxation of young clusters]{shukurea17}.
\item If the high mass slope of the CIMF is fixed to $x_2=1$, a satisfactory fit of the age distribution cannot be obtained. The basic assumption of a universal CIMF may need to be relaxed. The CIMF could depend on properties of the disc (gas fraction, stability) via a maximum cluster mass.
\end{itemize}

For a deeper understanding of cluster formation and evolution an extensive parameter study of the CIMF, the CFR, and the cluster mass evolution (instead of the cluster lifetime) is necessary. The theoretical predictions should then be compared to the 2-dimensional mass-age distribution. From the observational side, biases in terms of incompleteness and cluster mass determinations need to be understood in more detail.

\section{Summary and conclusions}\label{sec:conc}

In this study we constructed and investigated the age distributions of clusters using data from the all-sky survey of Galactic open clusters MWSC, which provides uniform and accurate ages as well as other relevant parameters like distances and reddenings. For assembling the distribution, we use a total of 2242 clusters located within the completeness radius of about 2.5 kpc from the Sun. Our sample is one order of magnitude larger than any previous samples used for age distribution analysis. Comparison with the literature shows that earlier results published in the 1980s-1990s strongly underestimate the fraction of evolved clusters with ages $\log t\gtrsim 8$. Recent studies, based on all-sky catalogues, agree more with our data, but still suffer from a lack of clusters older than about 1 Gyr. 

In order to consider radial variations in the age distribution, we build three radial sub-samples occupying different spatial locations within the completeness zone (the Inner, Local and Outer segments). They manifest general agreement of the distributions representing both the inter-arm space and two different spiral arms (Sagittarius and Perseus). The only prominent distinction is an enhanced fraction of old clusters in the Inner sample compared to the other two. This feature may be the manifestation of a higher cluster formation rate in the past in the inner disk. At the same time, it seems that the cluster formation histories of the Local and of the Outer sample were similar in the past.

We also compare two vertical sub-samples (the planar and high-altitude samples, which we associate with the thin- and thick-disk populations respecitvely) and find very different distributions. As expected, the thin-disk distribution agrees in general with the ``radial'' samples, though a deficiency of old ($t\gtrsim 1$ Gyr) clusters exists. In contrast, the thick-disk distribution is completely deprived of young clusters ($t< 250$ Myr), and to a large degree also of intermediate-age objects ($t< 1$ Gyr). Nevertheless, both distributions complement each other and together reproduce the total age distribution of disk clusters, and can be regarded as representatives of different populations having different formation histories.

With simple assumptions on the cluster formation history, the cluster initial mass function, and the cluster lifetime, we can reproduce the observed age distribution. The cluster formation rate and lifetime function are strongly degenerate, which prevents us from disentangling the different formation scenarios. In all cases the cluster formation rate is strongly declining with time, and the cluster initial mass function is very shallow at the high mass end.

\begin{acknowledgements}
This study was supported by Sonderforschungsbereich SFB 881 "The Milky Way System" (subproject B5) of the German Research Foundation (DFG) and by  Russian Foundation of Basic Research grant 16-52-12027. We acknowledge the use of the Simbad database, the VizieR Catalogue Service and other services operated at the CDS, France, and the WEBDA facility, operated at the Department of Theoretical Physics and Astrophysics of the Masaryk University. We thank the referee for comments and suggestions that helped us improve the paper.
\end{acknowledgements}

\bibliographystyle{aa}
%\bibliography{/home/tolya/bibdbs/clubib}
\bibliography{clubib}

\end{document}